\documentclass{aa}  
\usepackage{graphicx}
\usepackage{natbib}
\usepackage{epsf}
\usepackage{psfig}
\usepackage{lscape}
\usepackage{txfonts}
\usepackage{balance}
\begin{document}
   \title{Radio spectral study of the cluster of galaxies Abell 2255}


    \author{R.F.Pizzo\inst{1} and A.G. de Bruyn\inst{1,2}}

     \institute{Kapteyn Institute, Postbus 800, 9700 AV Groningen, The Netherlands \\ \email{pizzo@astro.rug.nl}
         \and
             ASTRON, Postbus 2, 7990 AA Dwingeloo, The Netherlands\\
                   \\ }


 
  \abstract{Spectral index studies of halos,
    relics, and radio galaxies provide useful information on their origin and
    connection with merger processes.}{We present WSRT multi-wavelength observations of the
    galaxy cluster Abell 2255 at 25 cm, 85 cm, and 2 m. The spectral index
    images allowed us to study the integrated spectrum of halo and relic and
    to investigate the physical properties of the Beaver head-tail radio
    galaxy belonging to the cluster.}{In the radio halo, the spectral index is
    steeper at the center and flatter at the locations of the radio filaments,
    clearly detected at 25 cm. In the relics, the spectral index flattens,
    moving away from the cluster center. For the Beaver radio galaxy, the
    spectrum severely steepens from the head towards the end of the tail,
    because of the energy losses suffered by the relativistic particles.

 In the 2 m map, which is the first high-sensitivity image presented in the
 literature at such a long wavelength, a new Mpc-size emission region is
 detected between the known radio halo and the NW relic. Not detecting this
 feature in the more sensitive 85 cm observations implies that it must have a
 very steep spectrum ($\alpha \leq -2.6$).}{The observational properties of
    the radio halo suggest that either we are looking at a superposition of
    different structures (filaments in the foreground plus real halo in the
    background) seen in projection across the cluster center or that the halo
    is intrinsically peculiar. The newly detected extended region to the NW of
    the halo could be considered as an asymmetric extension of the halo
    itself. However, since radio halos are known in the literature as
    structures showing a regular morphology, the new feature could represent
    the first example of steep Mpc-size diffuse structures (MDS), detected
    around clusters at very low frequencies.}


\authorrunning{Pizzo and de Bruyn}
\titlerunning{Radio spectral study of the cluster of galaxies Abell 2255}
\maketitle
%

\section{Introduction}
\label{introduction} 

Galaxy clusters, which contains up to a few thousand galaxies and considerable
amounts of gas, are the most massive gravitationally bound objects in the
universe. Cosmological simulations \citep{1995MNRAS.275...56N} show that they
are not static, but instead grow and still form at the present epoch, as the
result of several merger events \citep[e.g.][]{2002ASSL..272..253E}.

In an increasing number of massive, merging, and X-ray luminous galaxy
clusters, large diffuse radio sources associated with the ICM have been
detected. They represent the most spectacular aspect of cluster radio
emission, and they cannot be associated with any individual galaxy. These
diffuse synchrotron radio sources are characterized by a typical size of about
1 Mpc, low surface brightness ($\sim$ 1 $\mu$Jy arcsec$^{-2}$ at 20 cm), and
steep radio spectrum ($\alpha \leq - 1$, S($\nu$) $\propto$
$\nu^{\alpha}$). They are classified \citep{fergiov} as {\it radio halos}, if
located at the center of the cluster and not significantly polarized, and {\it
  radio relics}, if lying at the cluster periphery and showing high
polarization percentages ($\sim$ 10\%--30\%). The synchrotron nature of their
radio emission indicates cluster-wide magnetic fields of approximately 0.1--1
$\mu$G, and of a population of relativistic electrons with Lorentz factor
$\gamma \gg 1000$.

The knowledge of the physical conditions in halos and relics is important for
a comprehensive physical description of the ICM. Since they are related to
other cluster properties in the optical and X-ray domains, these extended
features are directly connected to the cluster history and
evolution. Moreover, they provide a significant test for several theories
concerning the origin of relativistic particles in the ICM and particle
propagation in astrophysical plasmas.

 The origin of halos and relics is still matter of debate. It is suggested
 that relics relics are related to shocks either by Fermi-I diffuse
 acceleration of ICM electrons \citep{ensslin,2004NewAR..48.1119K} or by
 adiabatic energization of the relativistic electrons confined in fossil radio
 plasma (``ghosts''), released by a former active radio galaxy
 \citep{2001A&A...366...26E}. In the case of radio halos, it is required that
 the electrons are either re-accelerated \citep[{\it primary models}:
   e.g.][]{1993MNRAS.263...31T,2001MNRAS.320..365B,2001ApJ...557..560P,2004MNRAS.350.1174B,2003ApJ...584..190F,2005MNRAS.357.1313C}
 or continuously injected over the entire cluster volume by hadronic
 collisions \citep[{\it secondary models}:
   e.g.][]{1980ApJ...239L..93D,1999APh....12..169B,2000A&A...362..151D}.

Spectral index studies of halos and relics provide important information on
their formation and their connection to cluster merger processes. Given their
large angular size and steep spectra, low-frequency radio observations are
suitable for such investigations. There are only a few clusters for which
spectral index maps of halos have been published: Coma
\citep{1993ApJ...406..399G}, A2163, A665 \citep{2004A&A...423..111F}, A3562
\citep{2005A&A...440..867G}, A2744, and A2219 \citep{2007A&A...467..943O}. In
agreement with primary models, in all these objects the spectral index maps
show a patchy structure with, in some cases, a steepening of the spectrum with
increasing distance from the center to the edge. The detailed spectral index
distribution is known for the relics of A3667 \citep{1997MNRAS.290..577R}, the
Coma cluster \citep{1991A&A...252..528G}, S753 \citep{2003AJ....125.1095S},
A2744 and A2219 \citep{2007A&A...467..943O}, and A2345 and A1240
\citep{2009A&A...494..429B}. Apart from one of the two relics of A2345, in all
these other examples the spectral index flattens going from the regions closer
to the cluster center to the outer rim. This is consistent with the presence
of electron re-acceleration in an expanding merger shock.

Other important members of clusters in the radio domain are the radio
galaxies. Their emission often extends well beyond the optical boundaries of
the radio galaxies, out to hundreds of kilo-parsec, and hence it is expected
that the ICM would affect their structure. This interaction is proved by the
existence of radio galaxies showing distorted structures (tailed radio
sources). Spectral studies of radio galaxies with different morphologies at
different frequencies provide an excellent opportunity to test the ageing of
the electrons, under some standard assumptions.

A2255 is a nearby \citep[z=0.0806,][]{struble} and rich cluster, which has
been studied in several bands.  {\it ROSAT} X-ray observations indicate that
A2255 has recently undergone a merger \citep{burns1995,fer,miller}.  Recent
{\it XMM-Newton} observations revealed temperature asymmetries of the ICM
suggesting that the merger happened $\sim$ 0.15 Gyr ago, probably along the
E-W direction \citep{sakelliou}. Optical studies of A2255 revealed the
presence of kinematic substructures in the form of several associated groups
\citep{yuan}. This result, together with the large ratio of velocity
dispersion to X-ray temperature \citep[6.3 keV;][]{horner} indicates a
non-relaxed system.\\ When studied at radio wavelengths, A2255 shows the
presence of a diffuse radio halo (located at the center of the cluster) and of
a relic (at the cluster periphery), together with a large number of embedded
head-tail radio galaxies \citep{har}.  High resolution radio observations at
21 cm shown that the radio halo has a rectangular shape and a surface
brightness increasing from the center to the edge \citep{fer}. At this
location 3 highly polarized ($\sim$ 20\%--40\%) bright filaments perpendicular
to each other are detected \citep{gov} . Sensitive low-frequency observations
carried out with the Westerbork synthesis radio telescope (WSRT) at 85 cm
proved the presence of a large number of low-surface brightness extended
features around the cluster, that are likely related to Large Scale Structure
(LSS) shocks \citep{2008A&A...481L..91P}. In this work we combine these data
observations at 25 cm and at 2m in order to study the spectral index
properties of the structures belonging to A2255.

This paper is organized as follows. Sect. \ref{introduction} describes the
main steps of the data reduction and discusses a few issues related to
low-frequency radio observations. In Sect. \ref{results}, we present the final
maps of A2255 at the three wavelengths and in Sect. \ref{spectralanalysis} we
show the spectral index analysis for the halo and the relics. In
Sect. \ref{beaver}, we analyze the physical properties of the Beaver radio
galaxy and we test the ageing processes of the radiating electrons along the
tail. We discuss the implications of our results in Sect. \ref{discussion} and
summarize our work in Sect. \ref{summary}.

Throughout this paper we assume a $\Lambda$ cosmology with $H_{0}$ = 71 km
s$^{-1}$\ Mpc$^{-1}$, $\Omega_{m}=0.3$, and $\Omega_{\Lambda}=0.7$. All
positions are given in J2000 coordinates. The resolutions are expressed in RA
x DEC. At the distance of A2255, $1^{\hbox{$\prime$}}$ corresponds to 90 kpc.


\section{Observations and data reduction}
\label{observations}

The observations were carried out with the WSRT in three broad frequency
ranges with central wavelengths of about 25 cm, 85 cm, and 2 m.  The array
consists of fourteen 25 m dishes on an east-west baseline and uses earth
rotation to fully synthesize the uv-plane. Ten of the telescopes are on fixed
mountings, 144 meters apart; the four remaining dishes are movable along two
rail tracks. The interferometer can observe in different configurations that
differ from each other in the distance between the last fixed telescope and
the first movable one (RT9--RTA). In the array, the baselines can extend from
36 m to 2.7 km. At the high declination of A2255 the array does not suffer
from shadowing.

The pointing center of the telescope, as well as the phase center of the
array, was directed towards ${\rm RA} = 17 ^{\rm h}13^{\rm m} 00^{\rm s}$,
${\rm Dec} = +64\hbox{$^{\circ}$ }07\hbox{$^{\prime}$ }59\hbox{$\arcsec$}$,
which is the center of A2255. The time sampling is 30 s in the 25 cm and 85 cm
datasets, and 10 s for the 2 m observations. This is generally sufficient to
sample the phase fluctuations of the ionosphere and to avoid the smearing of
sources at the outer edge of the field. During each observation, two pairs of
calibrators, one polarized and one unpolarized, have been observed for 30
minutes each. Table \ref{observationsdetails} summarizes the observational
parameters.

The technical details about the datasets will be discussed in the next sub
sections. Here we give a short overview about the main steps taken for the
data reduction.  The data were processed with the WSRT-tailored NEWSTAR
reduction package following mostly standard procedures: automatic interference
flagging, self calibration, fast Fourier transform imaging, and CLEAN
deconvolution \citep{1974A&AS...15..417H}. Further flagging based on the
residual data after self calibration and model subtraction was done after each
self calibration iteration.  An on-line Hamming taper was used to lower the
distant spectral side lobe level \citep{1978ieee...66...51H}. The final
analysis was done using only the odd channels, because each Hamming tapered
output channel is a linear combination of its direct neighbors and
itself.\\ At 85 cm and 2 m, the data were flux-calibrated using 3C295, for
which we adopted a flux of 63 Jy and 95 Jy at central frequencies of 346 MHz
and 148 MHz , respectively. At 25 cm the flux scales have been set using the
flux calibrator CTD93, for which we adopted a flux of 5.2 Jy at 1220 MHz. The
flux scale at 2 m is still uncertain to a level that we estimate of 5\% and
1\% at 25 cm and 85 cm. We computed the flux densities of the calibrators
across the entire bands assuming a spectral index of $\alpha = -0.5$ and
$\alpha = -0.6$ for CTD93 and 3C295 respectively (see Table
\ref{calibrators}).

\begin{table}
\caption{The flux calibrators used during the data reduction.}
\label{calibrators}
\smallskip
\begin{center}
{\small
\begin{tabular}{cccc}
\hline
\hline
\noalign{\smallskip}
Wavelength & name                 & Flux & $\alpha$\\
           &                      & (Jy)&          \\     
\noalign{\smallskip}
\hline
\noalign{\smallskip}
25 cm &             CTD93         & 5.2 &  --0.5   \\
\noalign{\smallskip}
\hline
\noalign{\smallskip}
85 cm &             3C295         & 63   &   --0.6 \\
\noalign{\smallskip}
\hline
\noalign{\smallskip}
2 m   &             3C295         & 95   &   --0.6 \\
\noalign{\smallskip}
\hline
\end{tabular}
}
\end{center}
\end{table}

\begin{table*}
\caption{Observations overview.}
\label{observationsdetails}
\smallskip
\begin{center}
{\small
\begin{tabular}{ccccccc}
\hline
\hline
\noalign{\smallskip}
Wavelength & Frequency range &   Calibrators   &Observation ID & RT9--RTA     & Start date (UTC) & End date (UTC)\\
           &    (MHz)        &                 &               &   (m)       &                  &               \\
\noalign{\smallskip}
\hline
\noalign{\smallskip}
           &                 &                 & 10702691 & 36 & 2007/06/18 17:00:15 & 2007/06/19 04:58:50\\
           &                 &   3C286, CTD93  & 10702898 & 54 & 2007/06/28 16:20:55 & 2007/06/29 04:19:25\\
25 cm      & 1159--1298 MHz  &   3C147, 3C138  & 10702958 & 72 & 2007/07/03 16:01:15 & 2007/07/04 03:59:45\\
           &                 &                 & 10703066 & 90 & 2007/07/10 15:33:45 & 2007/07/11 03:10:35\\
\noalign{\smallskip}
\hline
\noalign{\smallskip}
           &                 &                 & 10602227 & 36 & 2006/05/09 19:36:20 & 2006/05/10 07:35:20\\
           &                 &                 & 10602239 & 48 & 2006/05/10 19:32:20 & 2006/05/10 07:31:20\\
           &                 & DA240, 3C295,   & 10602259 & 60 & 2006/05/11 19:28:20 & 2006/05/12 07:27:20\\
85 cm      & 310--380 MHz    & PSR1937+21,     & 10602224 & 72 & 2006/05/16 19:08:40 & 2006/05/17 07:07:40\\
           &                 & 3C48            & 10602372 & 84 & 2006/05/19 18:57:00 & 2006/05/20 06:56:00\\
           &                 &                 & 10602340 & 96 & 2006/05/22 18:45:10 & 2006/05/23 06:44:10\\
\noalign{\smallskip}
\hline
\noalign{\smallskip}
           &                 &                 & 10702727 & 36 & 2007/06/21 16:48:15 & 2007/06/22 04:47:05\\
           &                 &                 & 10703226 & 48 & 2007/07/19 14:58:05 & 2007/07/20 02:56:55\\
           &                 & DA240, 3C295,   & 10703186 & 60 & 2007/07/17 15:05:55 & 2007/07/18 03:04:45\\
2 m        & 115--175 MHz    & PSR1937+21,     & 10703004 & 72 & 2007/07/05 15:53:15 & 2007/07/06 03:52:05\\
           &                 & 3C48            & 10703056 & 84 & 2007/07/09 15:37:25 & 2007/07/10 03:36:15\\
           &                 &                 & 10703099 & 96 & 2007/07/12 15:25:35 & 2007/07/13 03:24:25\\
\noalign{\smallskip}
\hline
\end{tabular}
}
\end{center}
\end{table*}

\subsection{\it 25 cm dataset}
\label{25cmdataset}

At 25 cm, we observed A2255 for $4 \times 12$ hours with 4 different
configurations of the WSRT. By stepping the 4 movable telescopes at
18m increments, from 36m to 90 m, we pushed the grating lobe due to
the regular 18 m baseline increment to a radius of $\sim$ 1
\hbox{$^{\circ}$ }.

At this wavelength, the receiving band is divided into 8 contiguous
slightly overlapping sub-bands of 20 MHz centered at 1169, 1186, 1203,
1220, 1237, 1254, 1271, and 1288 MHz. Each sub-band is covered by 64
channels in 4 cross-correlations to recover all Stokes parameters.

Because of limitations of the software in handling files larger than
2.15 GB, the data reduction was done for each frequency band
independently.  Thus, the original dataset has been split in 8
sub-datasets, each containing the 4$\times$12 hours of data, but just
in one frequency band.

Because of strong RFI, 2 out of the original 8 frequency sub-bands
have not been used for imaging. Moreover, the presence of a strong
off-axis source caused phase errors affecting the central field of the
final image. The peeling procedure that we used to solve for those
problems was successful only in the lower 3 frequency bands, where the
problematic source is brighter (see Sect.\ref{offaxiserrors}
). Therefore, the total amount of data that we could use for imaging
was $\sim$ 35\%.  At 25 cm, the resolution is $14 \arcsec \times 15
\arcsec $.

\subsection{Off-axis errors in 25 cm maps}
\label{offaxiserrors}

High dynamic range imaging is seriously limited by the phase stability
of the atmosphere (troposphere and ionosphere), which causes the
presence of spiky patterns surrounding the brightest off-axis sources
within the field of view of the interferometer. This pattern is due to
the instantaneous fan beam response of the WSRT, which rotates
clockwise from position angle $+90\hbox{$^{\circ}$}$ to
$+270\hbox{$^{\circ}$}$ during the 12 hours synthesis time.

The selfcal algorithm tries to minimize the time variations between data and
the model of the sky, applying only one averaged time-dependent correction in
the uv plane. If in the field of view there are bright off-axis sources that
seem to move during the observation, due to differential tropospheric
refraction or rapid ionospheric phase instabilities, the selfcal will not be
able to correct simultaneously for the phase errors associated with
them. Thus, the final map will show artefacts around the sources, as secondary
lobes, or in the source itself, which will appear distorted. The only way to
remove the off-axis errors in the maps is to ``peel'' each problematic source
out of the dataset, using its own direction-dependent selfcal
corrections. This technique has been recently developed to solve for the
ionospheric phase disturbances in low-frequency observations
\citep[e.g.][]{2004SPIE.5489..180C,2007mru..confE.101I}, but it can be also
applied at higher frequencies, when phase errors due to the differential
tropospheric effects are also present in high dynamic range images.

\begin{figure*}
\begin{center} 
\includegraphics[scale=.4,angle=270]{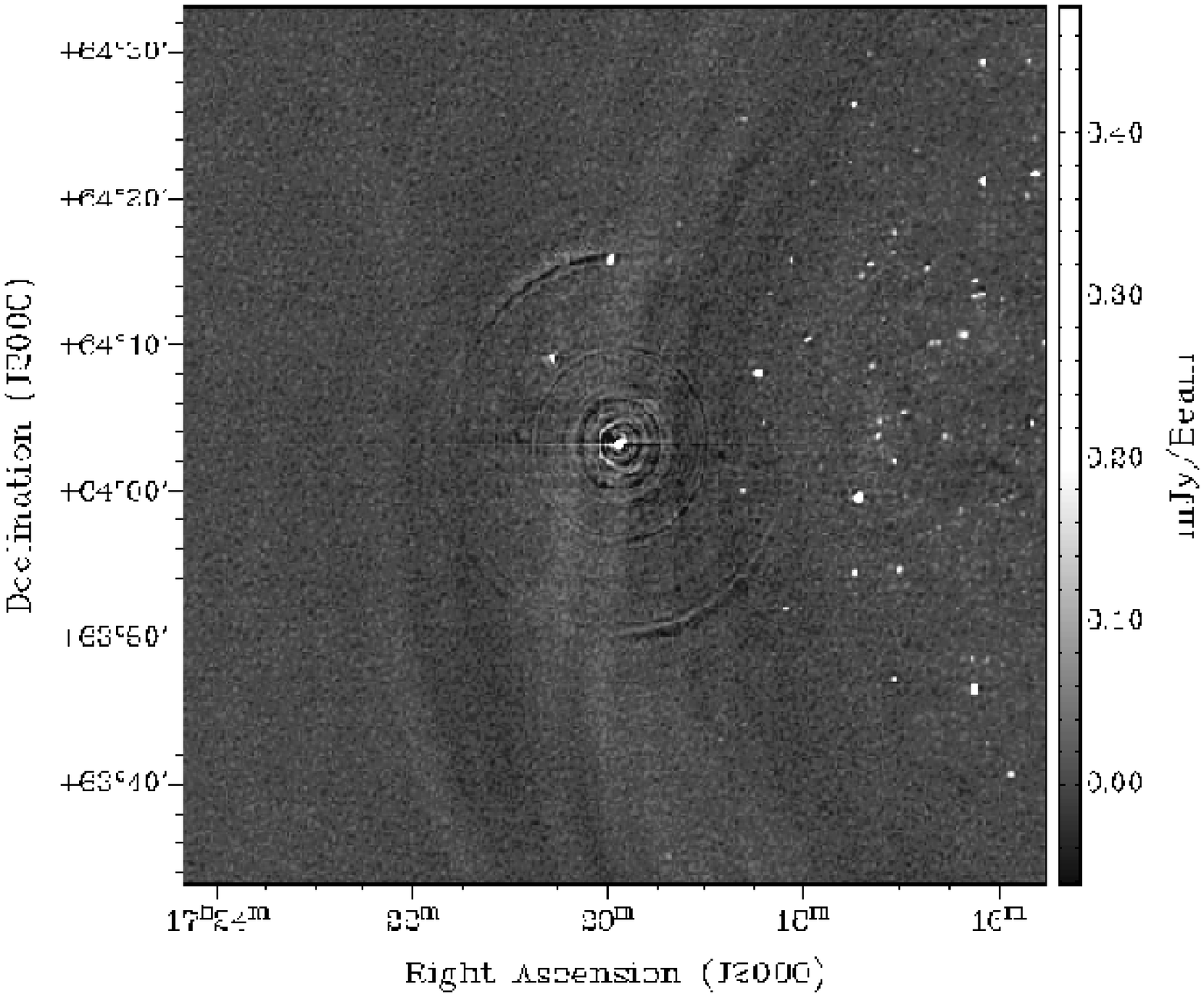}
\includegraphics[scale=.4,angle=270]{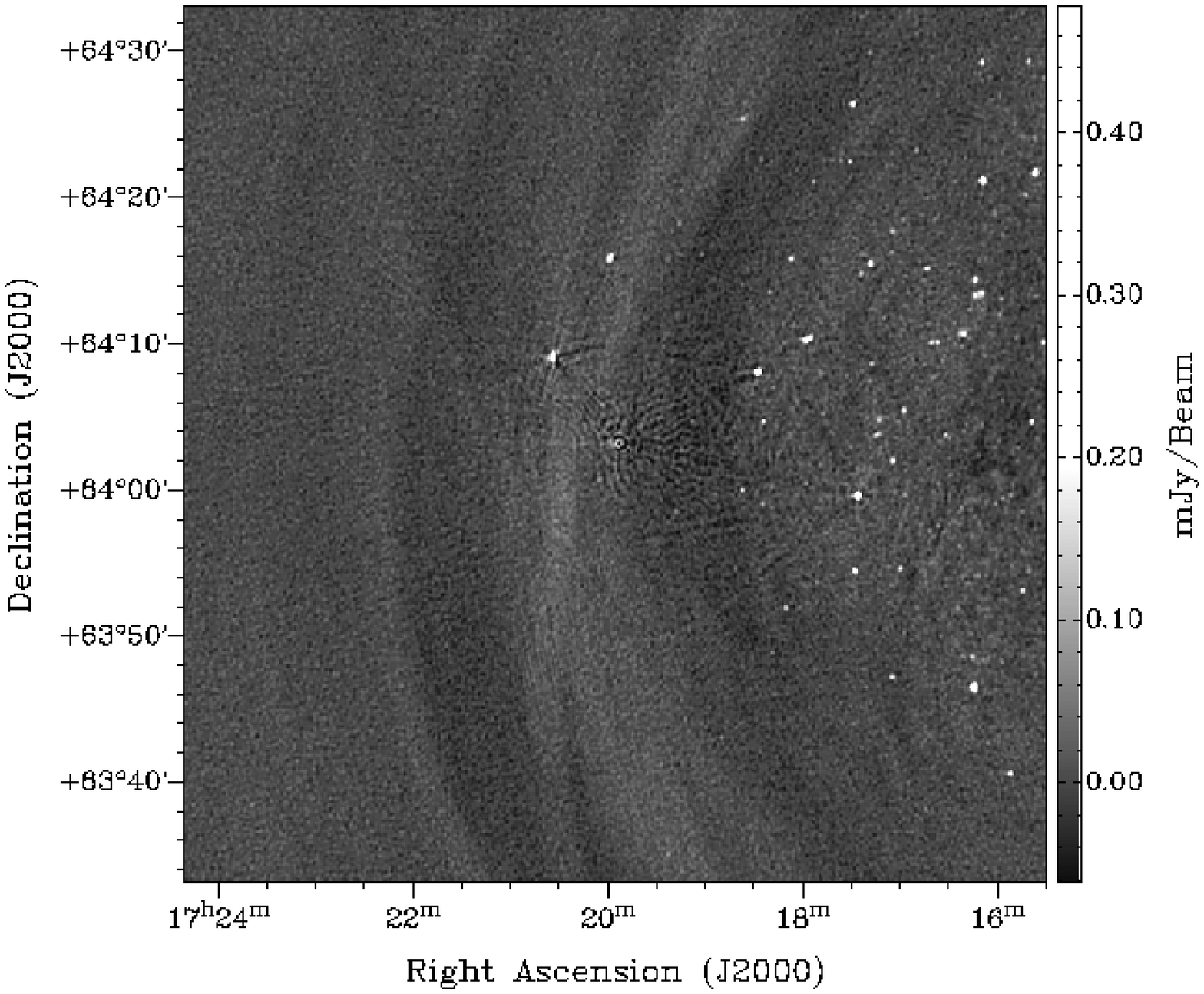}
\caption{Grey scale images of 4C +64.21 at 25 cm before (left panel) and after
  (right panel) applying the peeling procedure. The resolution is 14 $\arcsec
  \times$ 15 $\arcsec$. Left panel: the source shows phase-related errors,
  like ring structures and a negative and a positive spike along the east-west
  direction. Right panel: just a small residual is left at the source
  location.}
\label{peeling}
\end{center}
\end{figure*}

The peeling scheme involves self calibration on individual bright sources. It
produces phase corrections per array element for several viewing directions,
repeatedly using the following recipe:
\begin{itemize}
\item Subtraction of all but the brightest source from UV data, using the best
  model and calibration available;
\item several rounds of phase only self calibration and imaging on
the brightest source;
\item subtraction of the brightest source from the original UV data, using
  model and calibration from the previous step.
\end{itemize}
We applied this algorithm to the 25 cm dataset, which is the only one affected
by off axis errors that affect the analysis at the center of the maps. Also
our maps at 85 cm and at 2m showed off axis errors, but they were confined in
areas far away from the field center. Therefore no peeling procedure was
needed at these wavelengths.  At 25 cm the phase errors were associated with
the off axis source 4C +64.21, located at RA = 17 $^{\rm h}19^{\rm m} 59^{\rm
  s}$, Dec = $+64\hbox{$^{\circ}$ }04\hbox{$^{\prime}$
}40\hbox{$\arcsec$}$. In Fig. \ref{peeling}, we show the result of the
procedure. Without peeling, 4C +64.21 shows a distorted shape, with a positive
spike going towards east and a long negative spike directed towards the
opposite direction, where the center of the cluster is located. The source is
also surrounded by ring structures, which were left after the
deconvolution. After peeling, the source is basically removed. The flux below
which the sources have been subtracted for this procedure is $\sim$1
Jy. \\ The peeling procedure has been applied in each of the 6 frequency bands
we could use at 25 cm (Sect. \ref{25cmdataset}). Unfortunately, it turned out
to be satisfactory just for the three lower frequency bands, where the selfcal
corrections for the steep spectrum 4C +64.21 can be better determined, given
the higher signal to noise ratio of the data. Thus, we decided to use only
these 3 frequency bands to make the final image for the further analysis.

\subsection{\it 85 cm dataset}
\label{85cmdataset}

To fully image the primary beam, six array configurations were used for the
observations, with the four movable telescopes stepped at 12 m increments
(i.e. half the dish diameter) and the shortest spacing running from 36 m to 96
m. This provides continuous uv-coverage with interferometer baselines ranging
from 36 m to 2760 m.

The receiving band is covered by 8 sub-bands of 10 MHz centered at 315, 324,
332, 341, 350, 359, 367, and 376 MHz. Each sub-band is covered by 128 channels
in 4 cross-correlations to recover all Stokes parameters.  The 85 cm dataset
was split in 16 sub-datasets, each containing $6 \times 12$ hours of
observation, but in one frequency half-band only.

The imaging was done using 5 out of the 6 nights of observation, because of
the poor phase stability of the ionosphere during the third night (RT9--RTA =
60 m).  The total amount of flagged data was $\sim$ 25\% .  At 85 cm
wavelength, the resolution is $54 \arcsec \times 64 \arcsec$.

 \subsection{\it 2 m dataset (LFFE-band observations)}
\label{2mdataset}

At 2 m, A2255 was observed for 6 $\times$ 12 hours to fully image the primary
beam.\\ The LFFE (Low-Frequency Front End) at the WSRT is covered by 8
frequency sub-bands of 2.5 MHz centered at 116, 121, 129, 139, 141, 146, 156,
and 162 MHz. Each sub-band is divided in 512 channels.

Given the large size of the final dataset ($\sim$ 250 GB) and the limitations
in the maximum size of a NEWSTAR dataset, the observations have been split in
240 sub-datasets, each containing data for a single night of observation and
with a limited number of channels. Therefore, for a single spacing and for
each sub-band, we worked with 5 datasets. To produce the final map, we
processed the 240 sub datasets separately, self calibrating them with
different models and doing the imaging and the deconvolution steps 240 times.
We did not obtain a good image through the normal selfcal procedure. The best
image has been obtained using a simplified self calibration where we only
solved for the phase slope across the array.  In this case, only the phases
were self calibrated. At the end, the images were combined, weighting them for
the input number of visibilities.

Due to serious problems with the correlators of the WSRT during the night
between the 17th July and the 18th July 2007 (RT9--RTA = 60 m), the final data
reduction was done using 5 $\times$ 12 hours of observation. Moreover, given
serious ionospheric instabilities during the last night of observations
(RT9--RTA = 96 m), we decided to not include this 12 hours run in imaging.
The total amount of flagged data was $\sim$ 35\% .  At 2 m, the resolution is
$163 \arcsec \times 181 \arcsec$.

\subsection {A few issues related to low-frequency observations}
\label{afewissues}

The low-frequency radio sky is very bright and populated by strong radio
sources, such as Cas A, Cyg A, Vir A and Tau A, whose flux densities vary from
1000-10000 Jy. Even far from the field center, the (relatively) high distant
sidelobe levels of the primary beam (``only'' --30 to --40 dB) keeps these
sources very bright, giving rise to significant side lobes in the final
images. Solar flares can also affect the data, causing interferences in the
short baselines and/or grating rings in the imaged central field.

During the editing and imaging processes at low frequency, we had to take care
of Cas A, Cyg A and the Sun. Cas A lies at a projected distance of $\sim$
$40\hbox{$^{\circ}$}$ from the cluster, Cyg A at $\sim$ $30\hbox{$^{\circ}$}$
and the Sun at $\sim$ $85\hbox{$^{\circ}$}$. At 85 cm and at 2 m, we used
models of Cas A and Cyg A and we subtracted them from the data during
imaging. To remove the interferences generated by the Sun in the 85 cm
dataset, we flagged the short baselines in the hour angles in which the Sun
was above the horizon ($+30\hbox{$^{\circ}$}$ $\le$ HA $\le$
$+90\hbox{$^{\circ}$}$). At 2 m, the Sun was high in the sky for most part of
the observation. In this case, we produced different cleaning models of this
source for the different nights and we subtracted them during the final
imaging.

\subsection{Flux scale at 2 m wavelength}
\label{fluxscaleat2mwavelength}

The flux scale at low frequencies is not very accurately determined. Until a
more definitive flux scale is in place (this is being developed for LOFAR) we
use the radio source 3C295 as the primary WSRT flux calibrator at low
frequencies. We have adopted a flux density at 150 MHz of 95 Jy for 3C295 and
a power-law spectral index of --0.6 in the frequency range from 115--175
MHz. We believe this number to be accurate to about 5\%.

A major component of the system noise (receiver + sky) at low frequencies is
due to our Galaxy. The System Equivalent Flux Density (SEFD) of the telescopes
at 150 MHz is about 8000 Jy in the Galactic areas where 3C295 happens to be
located, but it rises to well over 10000 Jy in the Galactic plane.  Because
the WSRT receivers operate with an automatic gain control (AGC) system before
the analog-to-digital converter, it continuously measures the total power to
allow corrections for the variable input levels.  Unfortunately, most of the
time the total powers detectors (which integrate the power over the whole 2.5
MHz sub-band) are corrupted by RFI so we can not automatically correct the
correlation coefficients for the variations in system noise. The LFFE band is
full of mostly impulsive and narrow band RFI coming from airplanes,
satellites, and mobile users as well as electronic hardware within the
building which is located halfway the array.  At the high spectral resolution
(10 kHz) provided by the backend most of this RFI can be excised. However, the
total power data must be manually inspected for suitable stretches of power
level measurements.  These data form the basis for a manual correction of the
flux scale.

For A2255 the total power ratio between the cluster region and the 3C295 field
is 1.15 $\pm$ 0.05 at 141 MHz. Following the transfer of the complex gain
correction determined for 3C295 we have therefore applied an additional
correction of a factor 1.15 to the visibility data.

\begin{figure*}
\begin{center}
\includegraphics[scale=.3,angle=0]{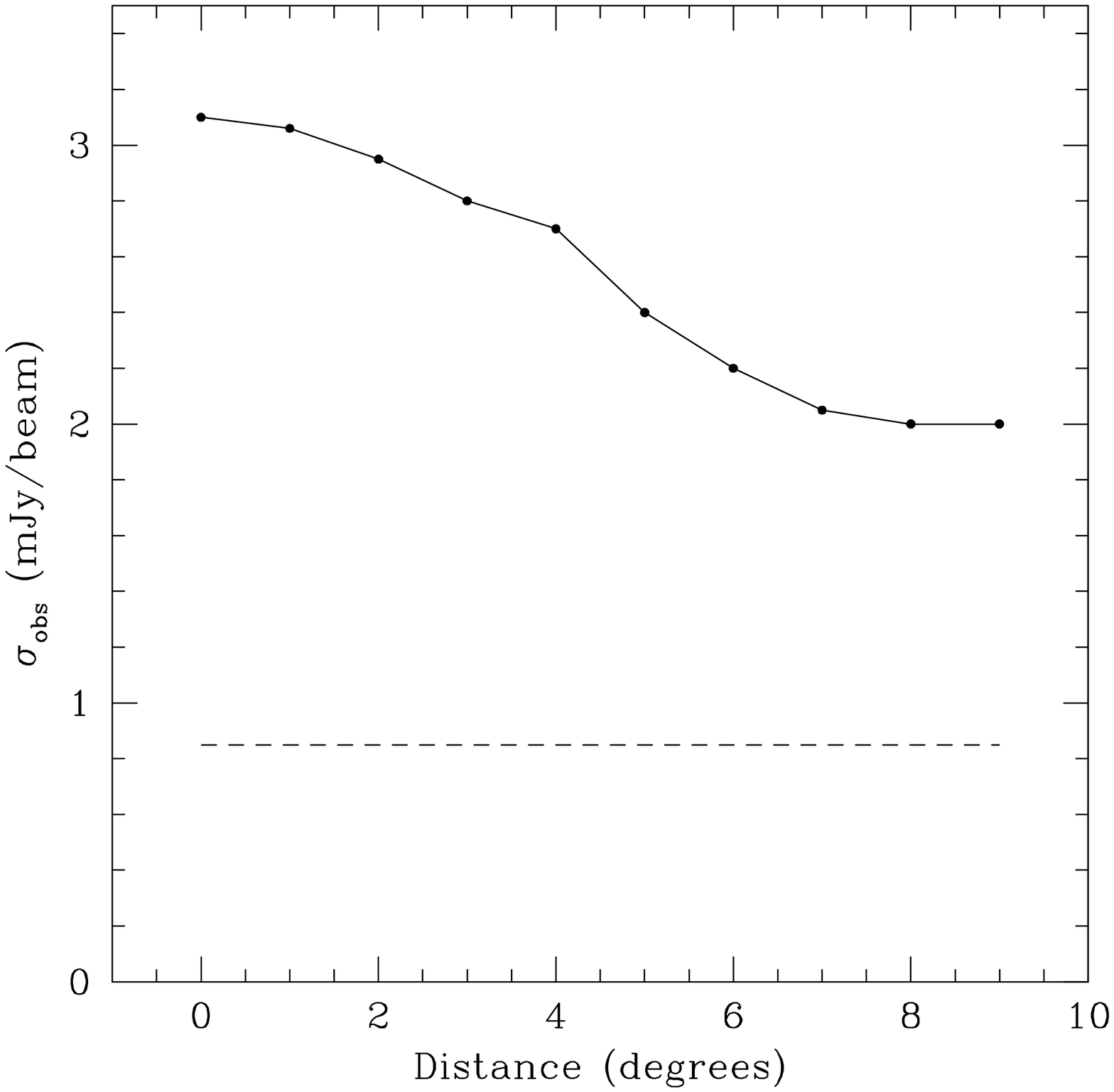}
\includegraphics[scale=.3,angle=0]{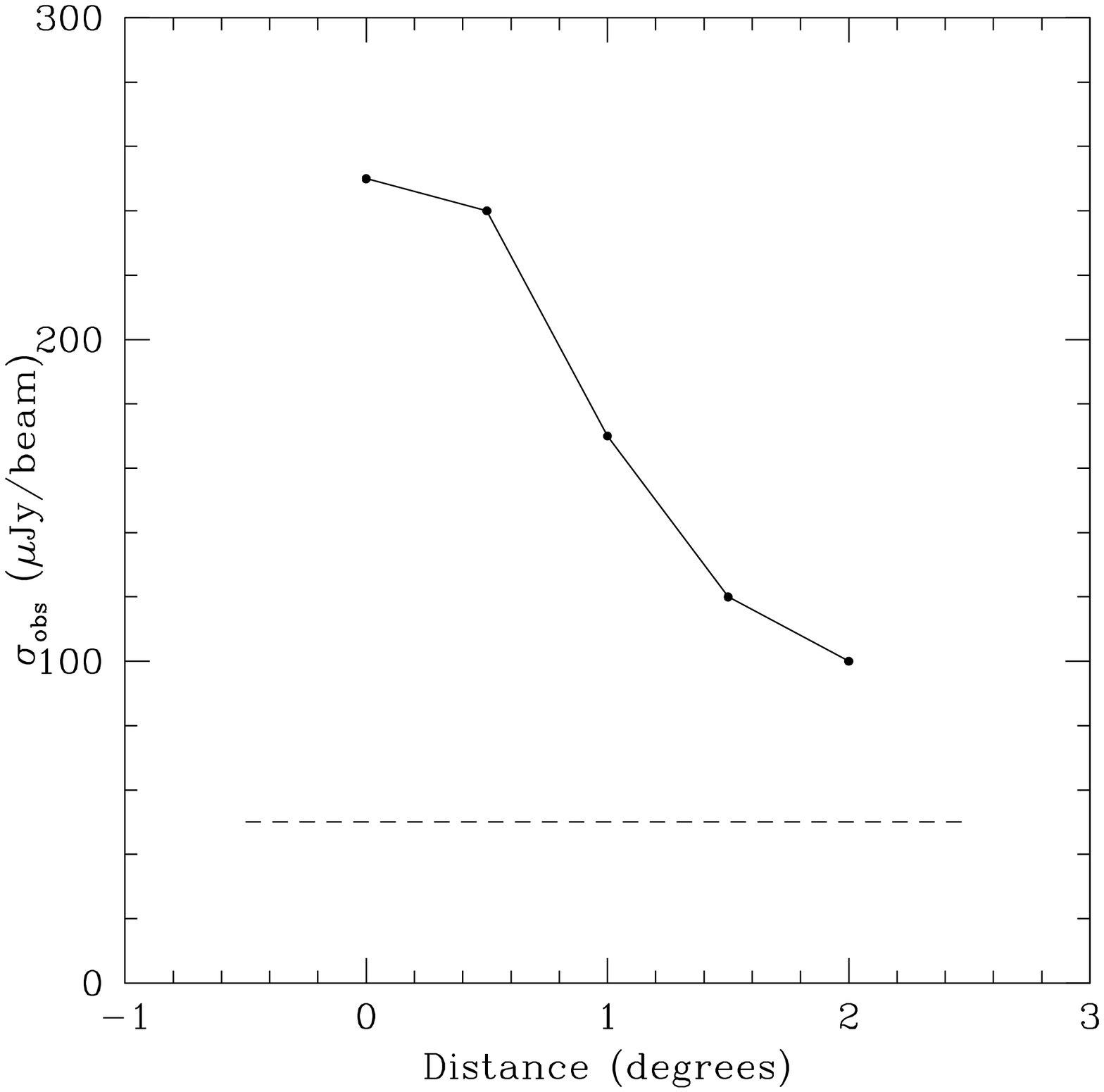}
\includegraphics[scale=.3,angle=0]{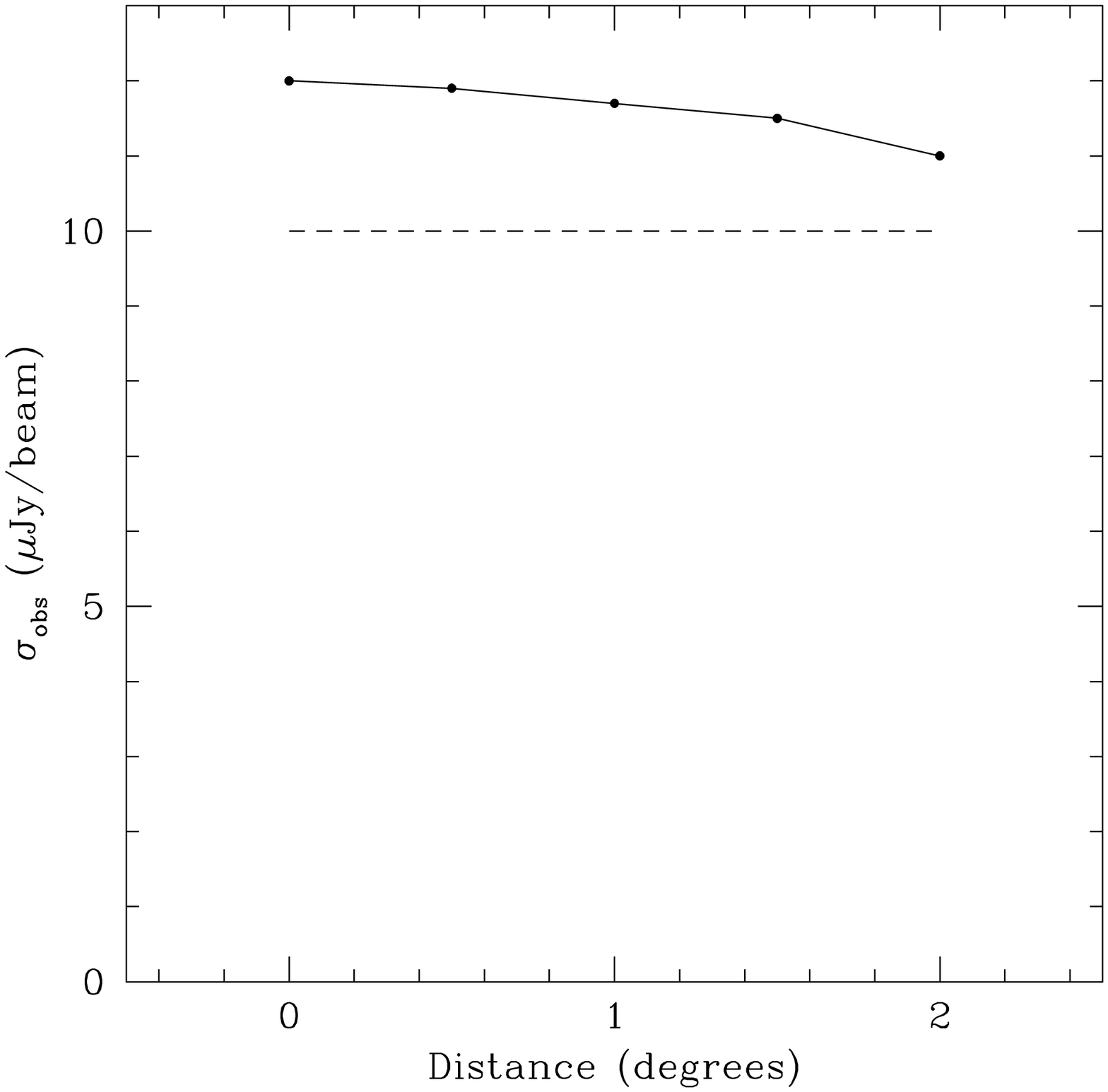}
\caption{Observed noise level uncorrected for the primary beam as a function
  of distance from the field center for the 2 m (left panel), 85 cm (middle
  panel), and 25 cm (right panel) maps. The dashed line represents the
  estimated thermal noise level, again uncorrected for the primary beam.}
\label{sigmaobs}
\end{center}
\end{figure*}

\begin{figure*}[!]
\begin{center}
\includegraphics[scale=0.60,angle=0]{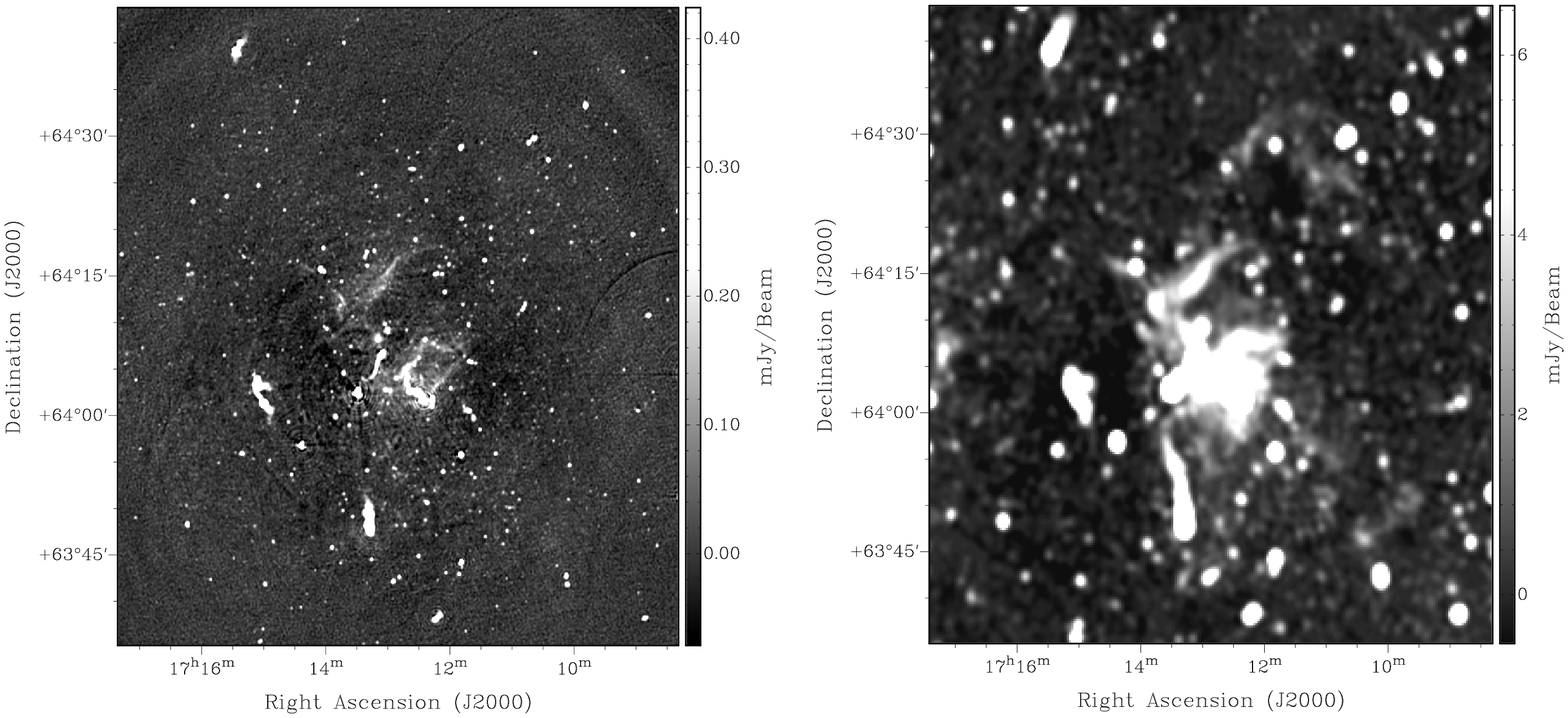}
\includegraphics[scale=0.40,angle=270]{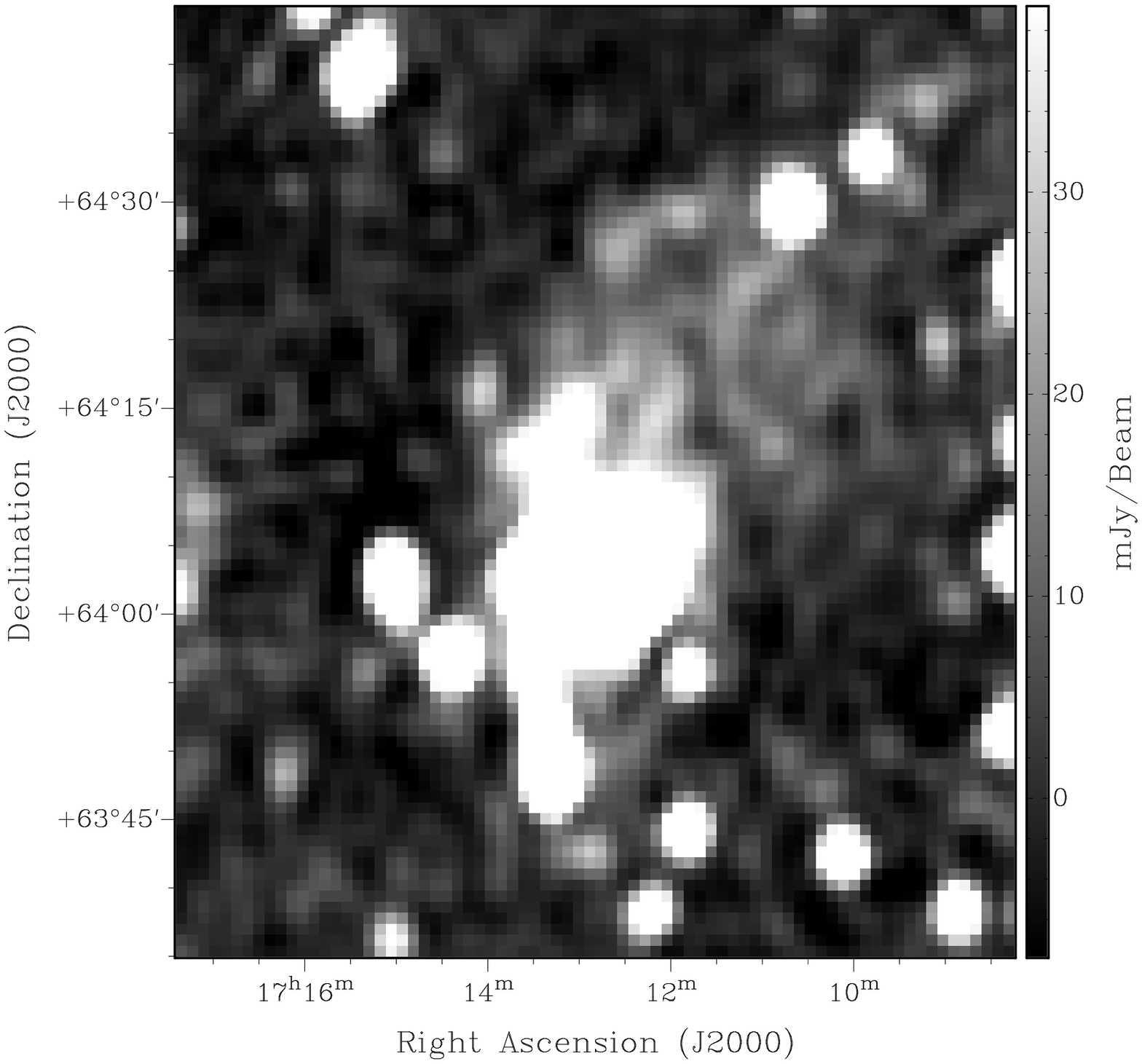}
\caption{Grey scale images of A2255 at 25 cm (top left panel), 85 cm (top
  right panel), and 2 m (bottom panel). All images cover the same area of the
  sky. The resolutions are 14 $\arcsec \times$ 15 $\arcsec$ (25 cm), 54
  $\arcsec \times$ 64 $\arcsec$ (at 85 cm) and 163 $\arcsec \times$ 181
  $\arcsec$ (at 2 m). The maps are not corrected for the primary beam.}
\label{A2255_25_85_2}
\end{center}
\end{figure*}

\begin{figure*}[!]
\begin{center}
\includegraphics[scale=0.65,angle=0]{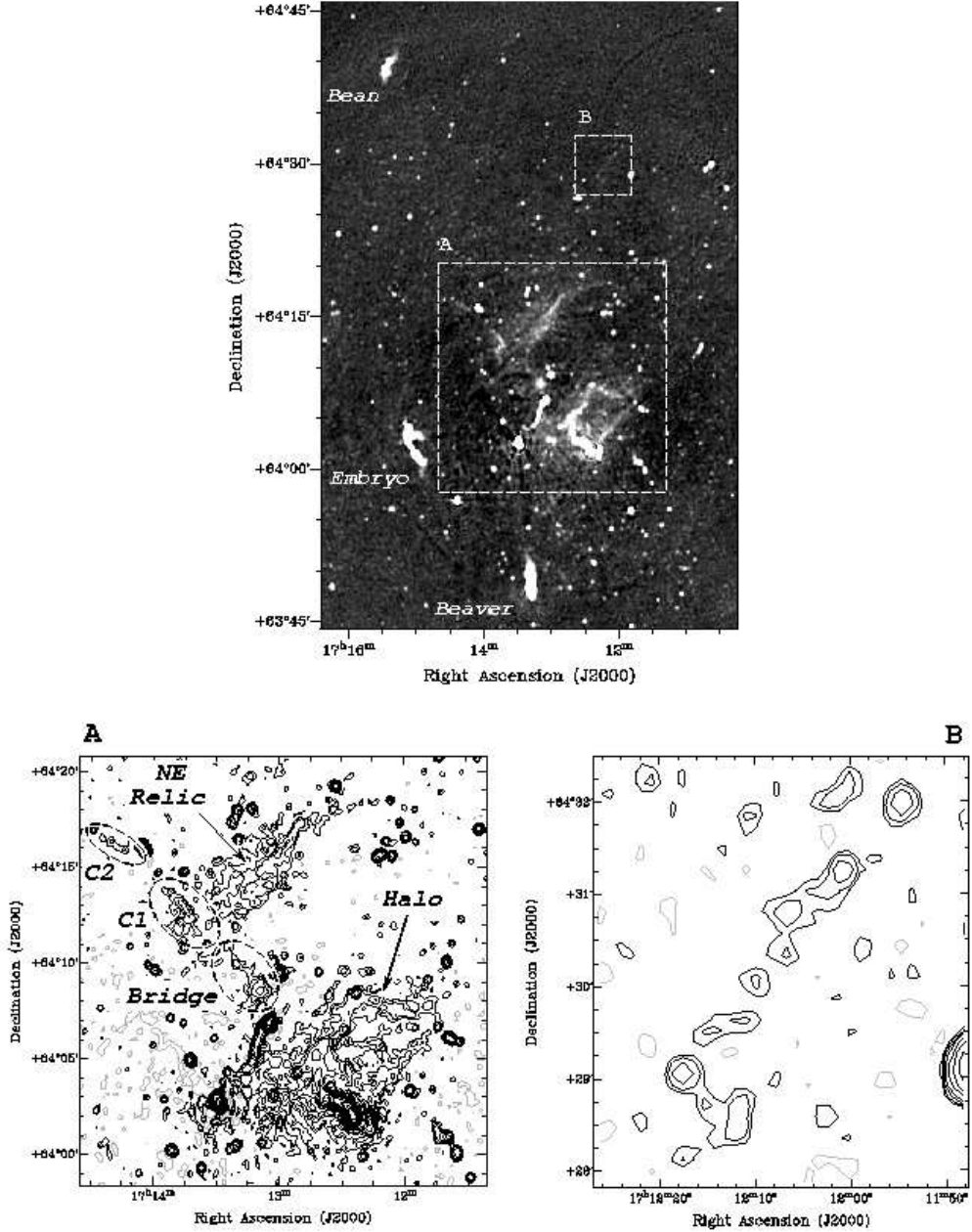}
\caption{Top panel: big field grey scale image of A2255 at 25 cm. The radio
  galaxies Bean, Embryo, and Beaver are visible, at large distance from the
  cluster center. The resolution is 14 $\arcsec \times$ 15 $\arcsec$.  The
  noise level at the edge of the fully imaged field is 11 $\mu$Jy/beam (see
  Fig. \ref{sigmaobs}). Bottom left panel: zoom into the central cluster
  region. We can distinguish halo, NE relic, bridge connection between them
  and the radio galaxies Goldfish, Double, The original TRG, and Sidekick. C1
  and C2 seem to be an extension of the radio bridge (see text). The contours
  are --0.03, 0.03, 0.06, 0.12, 0.24, 0.48, 0.96, 2, 4, 8, 16, 32 mJy/beam.
  Bottom right panel: zoom into the region of the cluster where the new NW
  relic has been detected at 85 cm (Sect. \ref{85cmmap}). At 25 cm, the only
  filament pointing towards the cluster center (NW1) is detected, but at very
  low level (1$\sigma$). Here the contours are --0.01 (grey), 0.01, 0.02,
  0.04, 0.06, 0.08, 0.16, 0.32, 0.64, 1.2, 2.4 mJy/beam. The maps are not
  corrected for the primary beam.}
\label{25cmfig}
\end{center}
\end{figure*}

\begin{figure*}[!]
\begin{center}
\includegraphics[scale=0.8,angle=0]{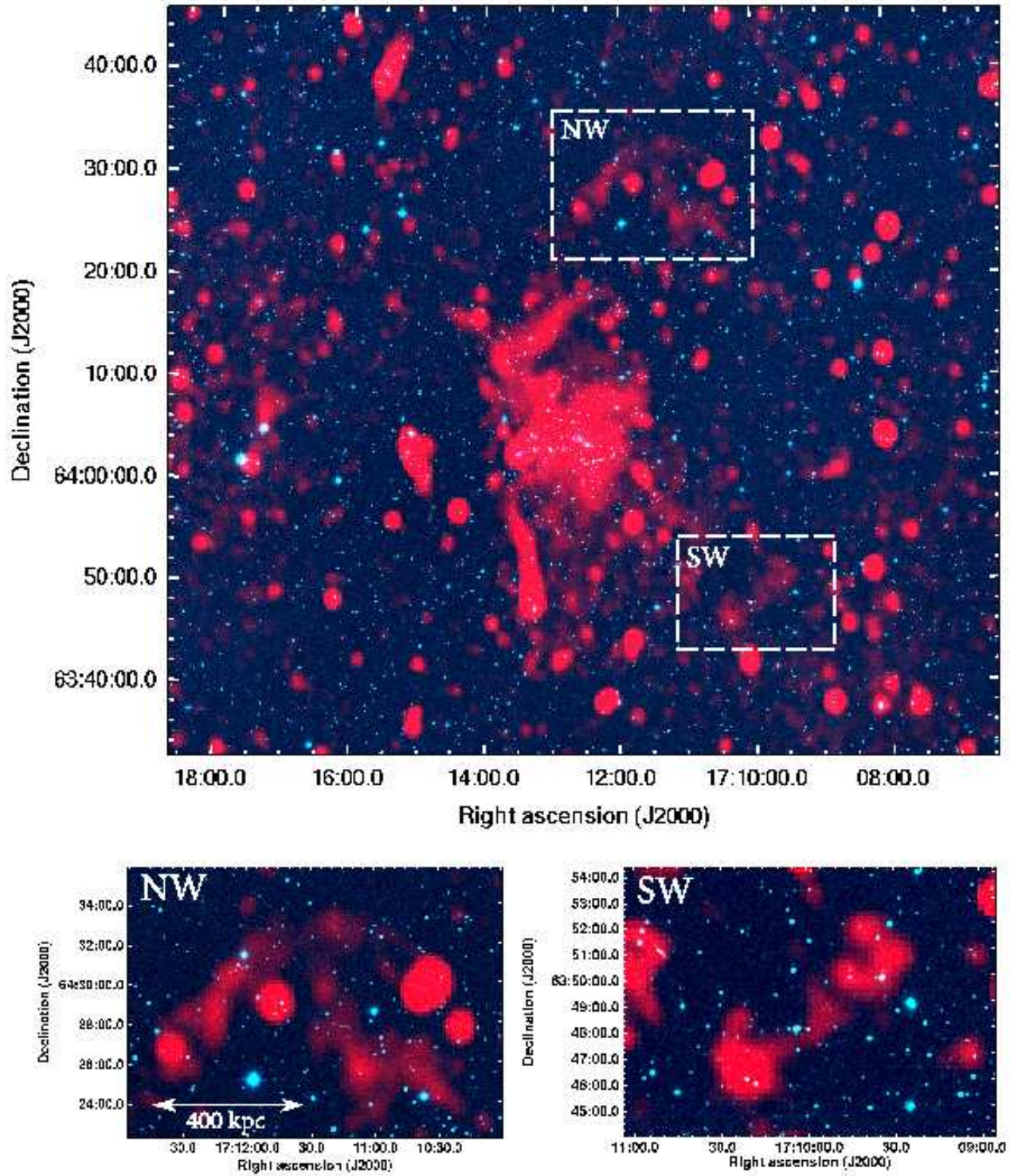}
\caption{Composite images of A2255 obtained from superposing the radio and
  optical images. The WSRT 85 cm radio map (in red) for the total field
  (central panel), NW relic (bottom left panel), and SW relic (bottom right
  panel) are shown overlaid on the red band Digitized Sky Survey image (blue). The radio image has a resolution of $54{\arcsec} \times
  64{\arcsec}$ and it is not corrected for the primary beam.}
\label{A2255_85CM_DSS_PAPER}
\end{center}
\end{figure*}

\begin{figure*}[!]
\begin{center}
\includegraphics[scale=0.82,angle=0]{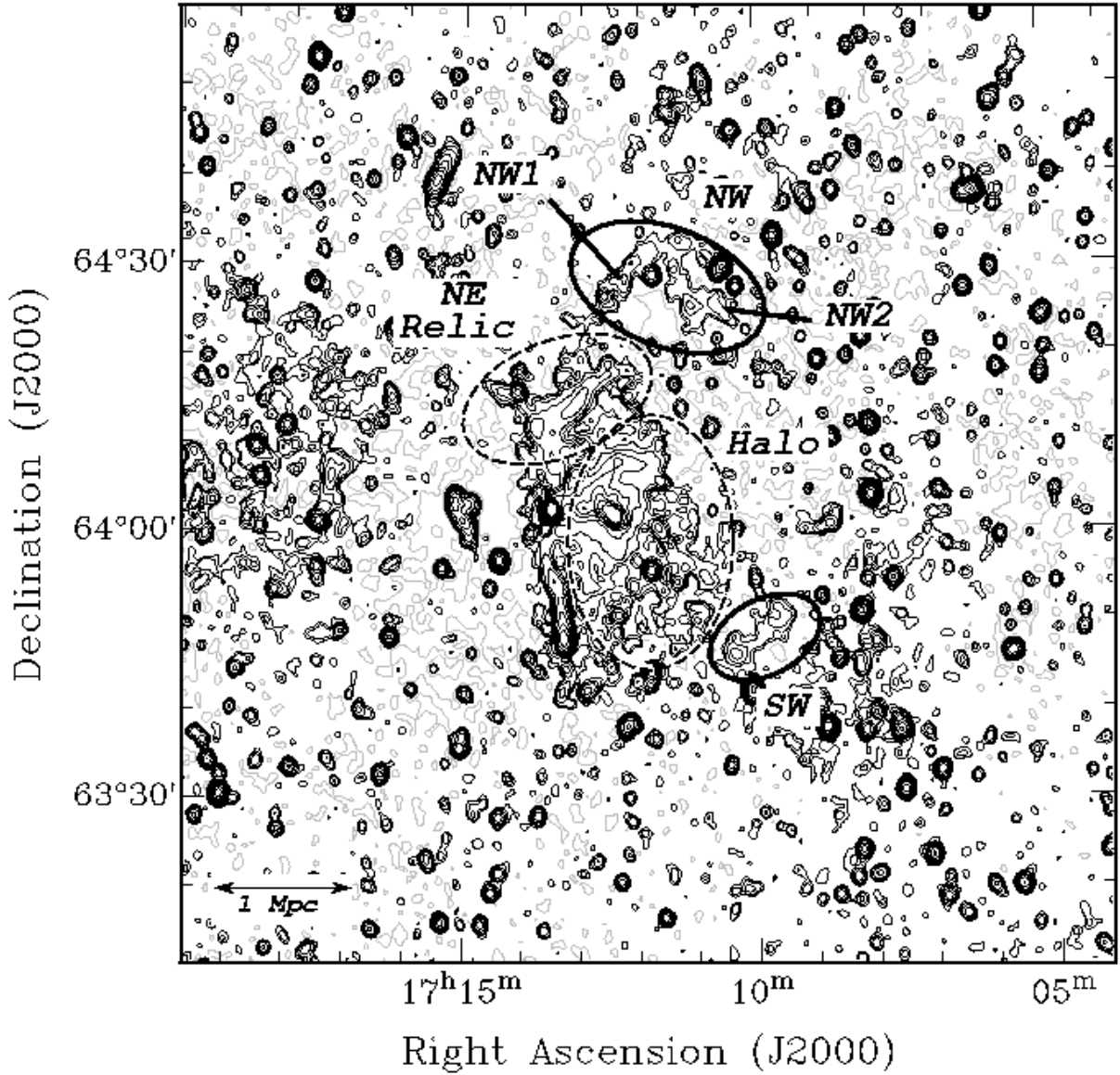}
\caption{Contour map of A2255 at 85 cm. The resolution is $54{\arcsec} \times
  64{\arcsec}$. The noise level at the edge of the fully imaged field is
  $\sim$ 0.1 mJy/beam (see Fig. \ref{sigmaobs}). The contours are --0.3
  (grey), 0.3, 0.6, 1.2, 2.4, 4.8, 9.6, 20, 40, 80, 160 mJy/beam. The dotted
  ellipses indicate the structures already known through previous studies, as
  the halo and the NE, while the solid ellipses refer to the newly detected
  relics. We refer to the text for a discussion about the positive features at
  the east side of the cluster center. The map is not corrected for the
  primary beam.}
\label{85cmfigure}
\end{center}
\end{figure*}

\begin{figure*}[!]
\begin{center}
\includegraphics[scale=.80,angle=0]{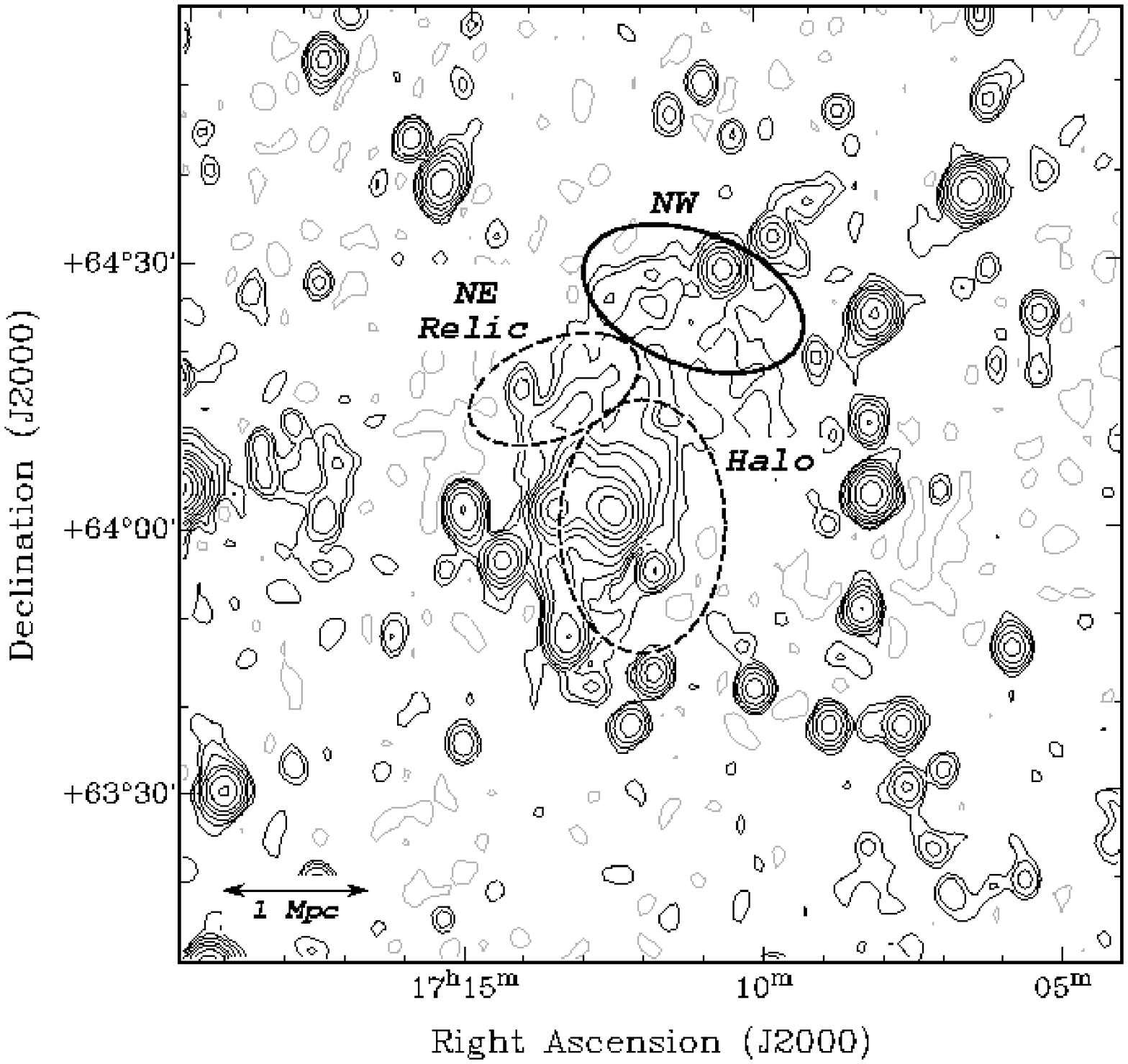}
\caption{Contour map of A2255 at 2 m. The resolution is $163{\arcsec} \times
  181{\arcsec}$. The noise level at the edge of the fully imaged field is
  $\sim$ 2 mJy/beam (see Fig. \ref{sigmaobs}). The contours are --0.007
  (grey), 0.007, 0.014, 0.028, 0.056, 0.1, 0.2, 0.4, 0.8, 1.6, 3.2
  Jy/beam. The radio halo is more extended towards NW than at 85 cm and is
  directly connected to the NW relic. The dotted lines indicate the structures
  already known through previous studies, as the halo and the NE, while the
  solid lines refer to the newly detected relics. The map is not corrected for
  the primary beam.}
\label{2mfigure}
\end{center}
\end{figure*}

\subsection{Errors and noise in the final images}
\label{errors}

Knowing the value of the error in an image is essential in determining the
reliability of the image and of the derived parameters. The observed errors
($\sigma_{obs}$) in a map mainly consist of:

\begin{itemize}
\item {the {\it thermal noise} ($\sigma_{th}$), which is due to the stochastic
  errors, coming from the Galaxy, receivers and other electronics used in the
  array;}
\item {the {\it confusion noise} ($\sigma_{conf}$), which actually consists of
  three contributions:
\begin{enumerate}
\item the ``normal'' sidelobe noise, which arises from the sum of all the
  sidelobes responses to the very large number of sources visible within the
  field of view. This noise can be minimized by deconvolution;
\item the ``classical'' source confusion, which is associated with the
number of sources within the same beam;
\item the ``error'' sidelobe noise, which results from calibration errors and
  non-isoplanaticity.
\end{enumerate}
}
\end{itemize}
It is important to appreciate the difference between thermal and confusion
noise. While the former is different every time we observe a source, the
latter is {\it always} the same for a given PSF, i.e. it repeats itself every
time we observe the same field.

In Table \ref{noises}, we list the estimated $\sigma_{th}$ and $\sigma_{conf}$
for the three observing wavelengths.  Reliable estimates of the thermal noise
have been obtained from polarization images (Q,U,V) and narrow spectral bands.
The confusion limit for WSRT at 21 cm was determined by
\citet{2000A&A...361L..41G} to be $\sim$ 5 $\mu$Jy. Because the PSF of the
WSRT depends on the declination, this value refers to Dec =
+6\hbox{$^{\circ}$}. At a different wavelength ($\lambda_2$) it is given by
\begin{equation}
\sigma_{\lambda_2} = \sigma_{\lambda_1} \left( \frac{\lambda_2}{\lambda_1} \right)^{2.75}\;\;, \nonumber
\end{equation}
where $\sigma_{\lambda_1}$ is the confusion limit at $\lambda_1$ = 21 cm and
the exponent (--2.75) takes into account the spectral index dependence
(--0.75) and the different synthesized beam (--2).\\ Assuming that
$\sigma_{th}$ and $\sigma_{conf}$ are uncorrelated, the observed noise in the
final map as a function of the distance from the center of the field is given
by :
\begin{equation}
\sigma_{obs}(r) = \sqrt{\sigma_{th} ^2 + \sigma_{conf} ^2(r)} \; .
\end{equation}
In Fig. \ref{sigmaobs}, we plot the observed noise levels, uncorrected for the
primary beam, as a function of the distance from the field center for the
final 2 m, 85 cm, and 25 cm full resolution maps. Because the classical
confusion noise will be attenuated by the primary beam, the observed noise
decreases from the center to the edges of the field, where it approaches the
thermal noise level.  This trend, more clear in the low-frequency maps, is a
problem when trying to assess the significance of the central extended
features in the contour maps. In principle, we should consider as detection
limit the noise observed at the center of the field and plot the contours
starting from, for example, 3 times this value. However, it is worth noting
that in this area we are confusion limited (this is a situation similar to
that found in deep optical images of galaxies, where the central fluctuations
are often dominated by the large number of stars within a resolution
element). Therefore, in order to show the significance of the features in our
final contour maps, we start plotting the contours from 3 times the noise
observed at the edge of the fully imaged field. To assess the significance of
the detected structures with respect to the background fluctuations, we advice
the reader to compare the contour maps at 25 cm, 85 cm, and 2 m with their
grey scale version in Fig. \ref{A2255_25_85_2}.

\subsection{Flux measurement uncertainties}
\label{fluxscaleuncertainties}

Because of the errors present in the final images, the flux and
spectral index estimates are affected by uncertainties.  The error
associated with the flux depends on:
\begin{itemize}
\item{the observed noise in the final maps ($\sigma_{obs}$);}
\item{the error due to the negative bowl, which arises around the
  extended structures because of the missing short spacings
  ($\sigma_{bowl}$). This severely affects the non full resolution
  maps only (see Sect \ref{makingthemaps}). In our case, its value was
  estimated by determining the mean value of the negative bowl around
  the central maximum in the antenna pattern};
\item{the uncertainty related with the flux of the calibrator
($\sigma_c$). This is a scale error which apply to the whole map.}
\end{itemize}
Under the assumption that these three uncertainties are uncorrelated,
the final error on the flux is
\begin{equation}
 \sigma_{F} = \sqrt{\sigma_c ^2 + \sigma_{obs} ^2 + \sigma_{bowl}
^2} \;\;\; .
\end{equation}
The uncertainty on the spectral index is given by 
\begin{equation}
\sigma_{\alpha} = \frac{1}{ln(\nu_2 / \nu_1)} \sqrt{\left( {\frac{\sigma_{F1}}{F_1}}\right ) ^2 + \left(  {\frac{\sigma_{F2}}{F_2}} \right )^2} \; ,
\end{equation} 
where $F_1 \pm \sigma_{F1}$ and $F_2 \pm \sigma_{F2}$ are the flux
densities and the corresponding errors at frequency $\nu_1$ and
$\nu_2$, respectively.

\begin{table*}
\caption{Parameters of the observations and of the final full
resolution maps.}
\label{noises}
\smallskip
\begin{center}
{\small
\begin{tabular}{ccccccc}
\hline
\hline
\noalign{\smallskip}
Wavelength           &   Full resolution   &   Largest detectable structure   &      ${\sigma_{\rm F}}^a$  &   ${\sigma_{th}}^b$   &   ${\sigma_{conf_{center}}} ^c$  &   ${\sigma_{obs_{center}}} ^d$  \\  
                     &    ${\arcsec}$      &          $^{\prime}$             &                         &     mJy/beam      &       mJy/beam             &           mJy/beam                    \\
\noalign{\smallskip}
\hline
\noalign{\smallskip}
2 m                  &    163$\times$ 181  &             191                  &         5\%             &        0.85       &            3               &            3.1                        \\
\noalign{\smallskip}
\hline
\noalign{\smallskip}
85 cm                &     54$\times$64    &              81                  &         1\%             &        0.05       &            0.25            &            0.25                        \\
\noalign{\smallskip}
\hline
\noalign{\smallskip}
25 cm                &     14$\times$15   &               24                  &         1\%             &        0.010      &            0.008           &            0.012                        \\
\noalign{\smallskip}
\hline
\noalign{\smallskip}
\noalign{\smallskip}
\multicolumn{7}{l}{$^a$ Flux scale uncertainty}\\
\multicolumn{7}{l}{$^b$ Thermal noise, constant over the maps}\\
\multicolumn{7}{l}{$^c$ Confusion noise}\\
\multicolumn{7}{l}{$^d$ Observed noise at the pointing position}
\end{tabular}
}
\end{center}
\end{table*}

\subsection{Making maps at 25 cm, 85 cm, and 2 m for a spectral index analysis}
\label{makingthemaps}

To make the final spectral index maps, we tapered the data to the same
low resolution of the 2 m dataset ($163\arcsec \times 181 \arcsec$),
and we cut the uv data with the same minimum uv coverage (144
$\lambda$), which is determined by the 25 cm data.\\
The missing short spacings in the low resolution 2 m and 85 cm maps
created a negative bowl around the central radio halo. To remove as
much as possible of the negative bowl, we cleaned very deep the halo
emission. At 2 m, the final map was obtained including in imaging only
the three frequency bands least affected by RFI.  Finally, the maps
have been corrected for the total power primary beam of the WSRT,
which can be approximated by
\begin{equation}
G(\nu,{\rm r}) = cos^6(c\nu {\rm r})~,
\end{equation} 
where c is a constant $\sim 0.064$, $\nu$ is the observing frequency
in MHz and r is the radius from the pointing center in degrees.


\section{Results}
\label{results}

A grey scale version of the final full resolution maps at 25 cm, 85 cm
and 2 m is shown in Fig. \ref{A2255_25_85_2}. In the following
subsections, we describe the features detected at each wavelength.

\subsection{25 cm map}
\label{25cmmap}

The 25 cm map is presented in Fig. \ref{25cmfig}. The image, which has
  a noise level of 11 $\mu$Jy at the edge of the field, shows the well
  known extended halo, located at the cluster center, the relic, that
  lies 10\hbox{$^{\prime}$} to the northeast from it, and 3 extended
  radio galaxies at a very large distance from the center of the
  cluster: the Embryo and the Beaver lie at $\sim$ 1.6 Mpc from the
  cluster center, while the Bean lies at more than 3.5 Mpc
  \citep[quoted names are taken from][]{har}. Since more relic
  features are detected at low frequency around A2255 (Sect. \ref
  {85cmmap}), from now on we will refer to the relic as NE
  (north-east) relic. Zooming into the central region of A2255, we
  notice the presence of 4 additional extended cluster radio galaxies:
  the Goldfish, the Double, the original TRG, and the Sidekick. The
  positions of the 7 radio galaxies are listed in Table
  \ref{radiogalaxies}. Each of them has an optical counterpart that
  belongs to the cluster \citep{2003AJ....125.2427M}.

The halo has a rectangular shape and shows a filamentary structure,
which is in agreement with previous 21 cm VLA observations
\citep{gov}. Given the high sensitivity of our new observations, halo
and NE relic look more extended and directly connected by a radio
bridge, which extends towards the north-east and seems associated with
two features; the former is located at ${\rm RA} = 17 ^{\rm h}13^{\rm m}
13^{\rm s}$, ${\rm Dec} = +64\hbox{$^{\circ}$
}12\hbox{$^{\prime}$}54\hbox{$\arcsec$}$ and belongs to the NE relic,
the latter lies at ${\rm RA} = 17 ^{\rm h}14^{\rm m} 18^{\rm s}$, ${\rm Dec} =
+64\hbox{$^{\circ}$ }16\hbox{$^{\prime}$}10\hbox{$\arcsec$}$. They
are labeled respectively C1 and C2 in Fig. \ref{25cmfig}. The physical
parameters of halo and NE relic are reported in Table
\ref{haloandrelic}.

We detect a very low surface brightness feature at location ${\rm RA} =
17^{\rm h} 12^{\rm m} 12^{\rm s}$, ${\rm Dec} = +64\hbox{$^{\circ}$
}30\hbox{$^{\prime}$}08\hbox{$\arcsec$}$, that is associated with one
of the two filaments of the NW relic detected at 85 cm
(Sect.\ref{85cmmap}). Since it lies in a pretty empty region of the
radio sky, a possible association of this feature with a collection of
point sources or with a central radio galaxy seems unlikely.

\begin{table}
\caption{The extended radio galaxies of A2255.}
\label{radiogalaxies}
\smallskip
\begin{center}
{\small
\begin{tabular}{ccc}
\hline
\hline
\noalign{\smallskip}
Name       &             RA (J2000)              &                       DEC (J2000)                              \\
           &                                     &                                                                 \\     
\noalign{\smallskip}
\hline
\noalign{\smallskip}
Bean       &   17 $^{\rm h}15^{\rm m}31^{\rm s}$ &   $+64\hbox{$^{\circ}$ }39\hbox{$^{\prime}$}28\hbox{$\arcsec$}$  \\
\noalign{\smallskip}
\hline
\noalign{\smallskip}
Beaver     &   17$^{\rm h}13^{\rm m} 19^{\rm s}$ &  $+63\hbox{$^{\circ}$}48\hbox{$^{\prime}$ }16\hbox{$\arcsec$}$   \\
\noalign{\smallskip}
\hline
\noalign{\smallskip}
Double     &   17 $^{\rm h}13^{\rm m} 28^{\rm s}$& $+64\hbox{$^{\circ}$ }02\hbox{$^{\prime}$}50\hbox{$\arcsec$}$    \\
\noalign{\smallskip}
\hline
\noalign{\smallskip}
Embryo     &   17 $^{\rm h}15^{\rm m}05^{\rm s}$ & $+64\hbox{$^{\circ}$}03\hbox{$^{\prime}$}42\hbox{$\arcsec$}$     \\
\noalign{\smallskip}
\hline
\noalign{\smallskip}
Goldfish   &   17 $^{\rm h}13^{\rm m} 04^{\rm s}$& $+64\hbox{$^{\circ}$ }06\hbox{$^{\prime}$}56\hbox{$\arcsec$}$    \\
\noalign{\smallskip}
\hline
\noalign{\smallskip}
TRG        &   17 $^{\rm h}12^{\rm m} 23^{\rm s}$& $+64\hbox{$^{\circ}$ }01\hbox{$^{\prime}$}46\hbox{$\arcsec$}$    \\
\noalign{\smallskip}
\hline
\noalign{\smallskip}
Sidekick   &   17 $^{\rm h}12^{\rm m} 16^{\rm s}$& $+64\hbox{$^{\circ}$ }02\hbox{$^{\prime}$}14\hbox{$\arcsec$}$    \\
\noalign{\smallskip}
\hline
\end{tabular}
}
\end{center}
\end{table}

\subsection{85 cm map}
\label{85cmmap}

The 85 cm map of A2255 is presented in Fig. \ref{85cmfigure}. With
noise levels ranging between 0.08 mJy/beam to 0.25 mJy/beam, limited
by classical confusion noise (see Sect. \ref{errors} and Table
\ref{noises}), it improves over previous imaging at close wavelength
\citep{fer} by a factor of 20. The overlay of the radio map with the
red band Digitized Sky Survey (DSS) optical image is presented in
Fig. \ref{A2255_85CM_DSS_PAPER}. This image clearly shows the
extension of the radio emission compared to the optical galaxies.

 The halo, the NE relic, and the radio galaxies belonging to the
 cluster are detected. The central radio halo looks much more complex
 than in previous images at the same frequency \citep{fer} and it is
 more extended than at 25 cm, in particular towards the S and SW. We
 notice that the southern region of the halo is directly connected, in
 projection, to the tail of the Beaver radio galaxy, which has doubled
 its length to almost 1 Mpc between 25 cm and 85 cm. Feature C2 in the
 25 cm map (see Fig. \ref{25cmfig}) is now more prominent and it looks
 directly associated to the NE relic.  The physical parameters of halo
 and NE relic at 85 cm are listed in Table \ref{haloandrelic}.

The high sensitivity of our observations allow us to detect two new
extended features at a projected distance of 2 Mpc from the cluster
center. The new ``relics'' are located NW and SW of the center of the
cluster and previous 21 cm images of A2255 revealed that they are
genuine features and not a collection of discrete sources
\citep{2008ASPC..386..368P}. From now on, we will call them the NW
(north-west) and SW (south-west) relic, respectively.  They have
different shapes. The SW relic appears like a filament of about
8$\hbox{$^{\prime}$ }$ in length and 2$\hbox{$^{\prime}$ }$ in
width. It has the same orientation of the known NE relic, but is
located on the opposite side from the cluster center and at a double
distance from it. The NW relic has a more complex morphology. We can
distinguish 2 filaments, labeled NW1 and NW2 in Fig. \ref{85cmfigure}.
NW1 points towards the cluster center and has a length of $\sim$
8$\hbox{$^{\prime}$ }$ $\times$ 1$\hbox{$^{\prime}$ }$, while NW2 is
$\sim$ 13$\hbox{$^{\prime}$}$ $\times$ 1$\hbox{$^{\prime}$}$
and is perpendicular to NW1. The SW and NW relics have integrated flux
densities of $\sim$ 17 mJy and $\sim$ 61 mJy, respectively.  The
physical properties of the newly detected structures and their origin
have been investigated by \citet{2008A&A...481L..91P}, who suggested a
connection with LSS shocks.\\ Other extended diffuse low surface
brightness features are detected to the east (${\rm RA} = 17 ^{\rm h}16^{\rm
m} 57^{\rm s}$, ${\rm Dec} = +64\hbox{$^{\circ}$
}18\hbox{$^{\prime}$}45\hbox{$\arcsec$}$) and to the west (${\rm RA} = 17^{\rm
  h}07^{\rm m} 08^{\rm s}$, ${\rm Dec} = +63\hbox{$^{\circ}$
}59\hbox{$^{\prime}$}36\hbox{$\arcsec$}$) of the cluster center. Their
nature is still unclear.  Moreover, the map shows positive and
negative fluctuations on a scale of 0.5 to 1 degree, which are likely
due to our Galaxy.

\subsection{2 m map}
\label{2mmap}

The 2 m map is shown in Fig. \ref{2mfigure}. The noise ranges between
2 mJy and 3 mJy and is limited by classical confusion noise in the
inner part of the map (see Sect. \ref{errors} and Table \ref{noises}).

\begin{figure}
\begin{center}
\includegraphics[scale=.42,angle=0]{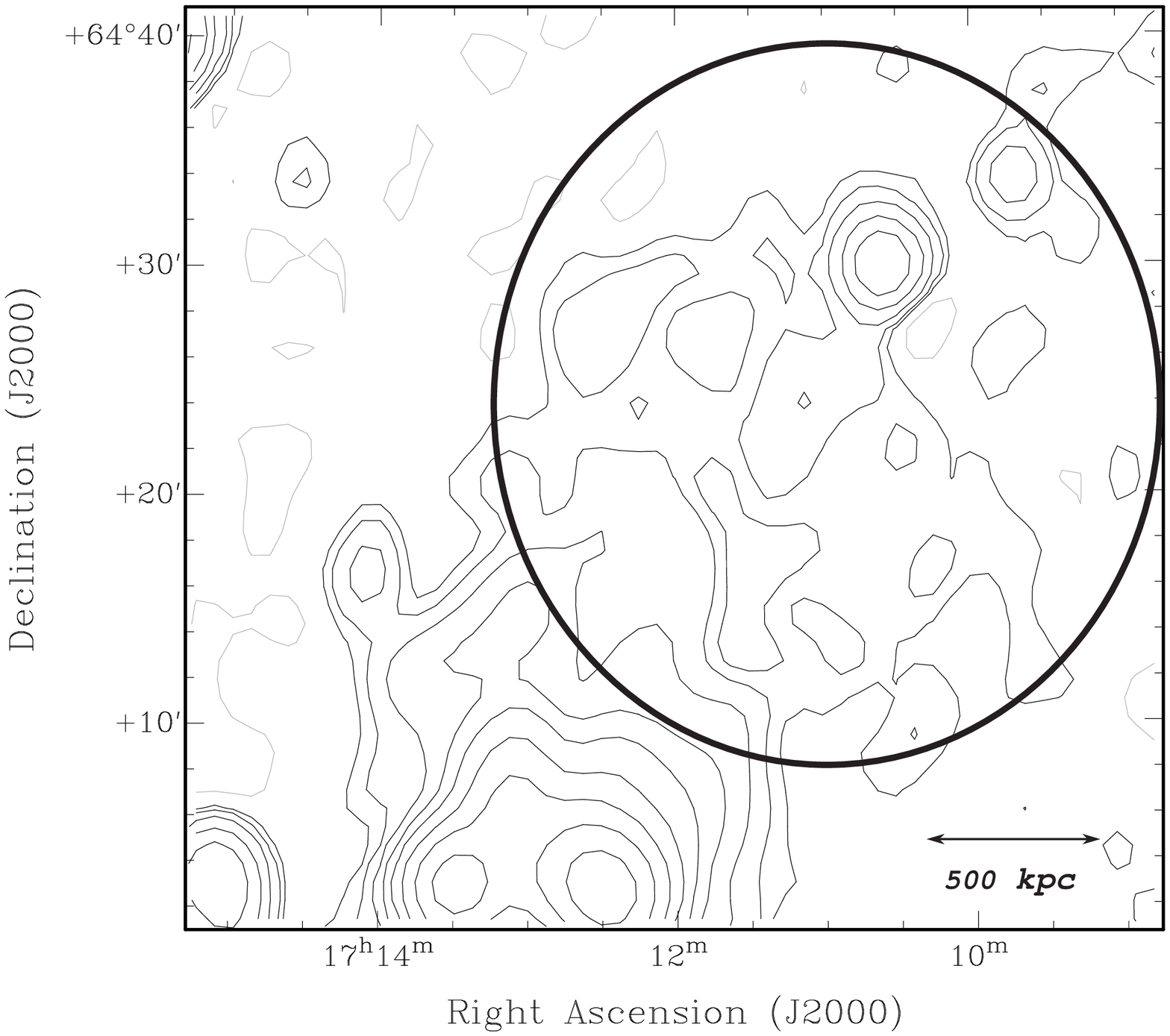}
\caption{Map at 2 m of the new extended emission region to the NW of
the halo of A2255 after the subtraction of the model of point sources
detected at 25 cm (see text). The result suggests that the new feature
is genuine and not due to a collection of point sources. The
resolution of the map is $163{\arcsec} \times 181{\arcsec}$. The
contours are --0.007 (grey), 0.007, 0.014 0.028 0.056, 0.1, 0.2, 0.4,
0.8 1.6 3.2 Jy/beam. The map is not corrected for the primary
beam. The circle represents the area within which the point sources
have been subtracted.}
\label{2mfiguresubtraction}
\end{center}
\end{figure}

Around the central radio halo, the known radio galaxies belonging to the
cluster are still detected. The diffuse emission associated with A2255 seems
to be rather complex.  The radio halo is extended towards NW and is connected
to the NW relic. The largest detectable structure in our 85 cm observations is
$\sim$ 1$\hbox{$^{\circ}$ }$, which means that non detecting this extended
feature at 85 cm cannot be due to uv plane coverage issues. Instead, it
supports the more likely hypothesis that the new emission region is a steep
spectrum feature, whose nature is different from the NW relic. We note that
point sources might make an important contribution to the new extended
emission feature. To test its nature, we removed the contribution of point
sources detected at 25 cm in this area, assuming a nominal spectral index of
$\alpha = -0.7$. The total flux subtracted in the region is 635 mJy. The
result, shown in Fig. \ref{2mfiguresubtraction}, confirms the genuine diffuse
nature of the feature. The source at location ${\rm RA} = 17^{\rm h}10^{\rm m}
37^{\rm s}$, ${\rm Dec} =
+64^{\circ}30^{\prime}24{\arcsec}$ is due to a blend of 2
point sources, as can be seen at higher frequency
(Fig. \ref{85cmfigure}). Given their steep spectrum ($\alpha \sim -1$), they
are still visible after the subtraction.  We computed the upper limit for the
spectral index of this feature using maps at 2 m and at 85 cm restored with
the same resolution ($163{\arcsec} \times 181{\arcsec}$), covering the same uv
range and in which we subtracted the point sources detected at 25 cm. In the 2
m map, the feature has a peak brightness of 28 mJy/beam, while in the 85 cm
map we can only give an upper limit, considering 3 times the noise of this map
($\sigma_{obs}$ = 1 mJy/beam), i.e. 3 mJy/beam. As a result, we obtain that
the newly detected feature should have a spectrum steeper than --2.6.

Because of the low resolution of the image, the NE relic seems to be
embedded in the halo emission.  The feature associated with it,
already detected at 85 cm and tentatively at 25 cm (C2 in
Fig. \ref{25cmfig}, bottom left panel), is now more
prominent. However, we notice that, in this case, part of it could be
due to the radio source located at ${\rm RA}= 17^{\rm h}14^{\rm m} 04^{\rm
s}$, ${\rm Dec}=+64^{\circ}16^{\prime}10{\arcsec}$, clearly detected at both 25
cm and 85 cm.\\ The physical parameters of halo and NE relic at 2 m
are listed in Table \ref{haloandrelic}. We note that it is difficult
to determine the real size and the borders of the halo and the NE
relic, therefore the integrated flux densities reported in the table
also reflect this uncertainty. The angular sizes of the two structures
are assumed to be the same than at 85 cm.

The NW relic is visible at 2 m, while the SW one, which has a surface
brightness lower than the NW relic at 85 cm, is not detected ($S_{SW
relic} <$ 6 mJy). This is mainly due to the confusion limit in the
central area of the map, where the noise level is 3 mJy.

Other extended features are detected to the E, NW and SW of the
cluster center and at very large projected distance from it. The first
one is located at ${\rm RA} = 17^{\rm h}17^{\rm m} 25^{\rm s}$, ${\rm Dec} =
 +64\hbox{$^{\circ}$ }05\hbox{$^{\prime}$}15\hbox{$\arcsec$}$, the
second one at ${\rm RA} = 17^{\rm h}09^{\rm m} 29^{\rm s}$, ${\rm Dec} = +64\hbox{$^{\circ}$ }36\hbox{$^{\prime}$}44\hbox{$\arcsec$}$ and the
third one at ${\rm RA} = 17^{\rm h}06^{\rm m} 10^{\rm s}$, ${\rm Dec} = +63\hbox{$^{\circ}$ }18\hbox{$^{\prime}$}42\hbox{$\arcsec$}$.
Comparing the 2 m map with the full resolution 85 cm and 25 cm images,
it is evident that in this case we are dealing with unresolved point
sources.

\begin{figure*}[!]
\begin{center}
\includegraphics[scale=.57,angle=270]{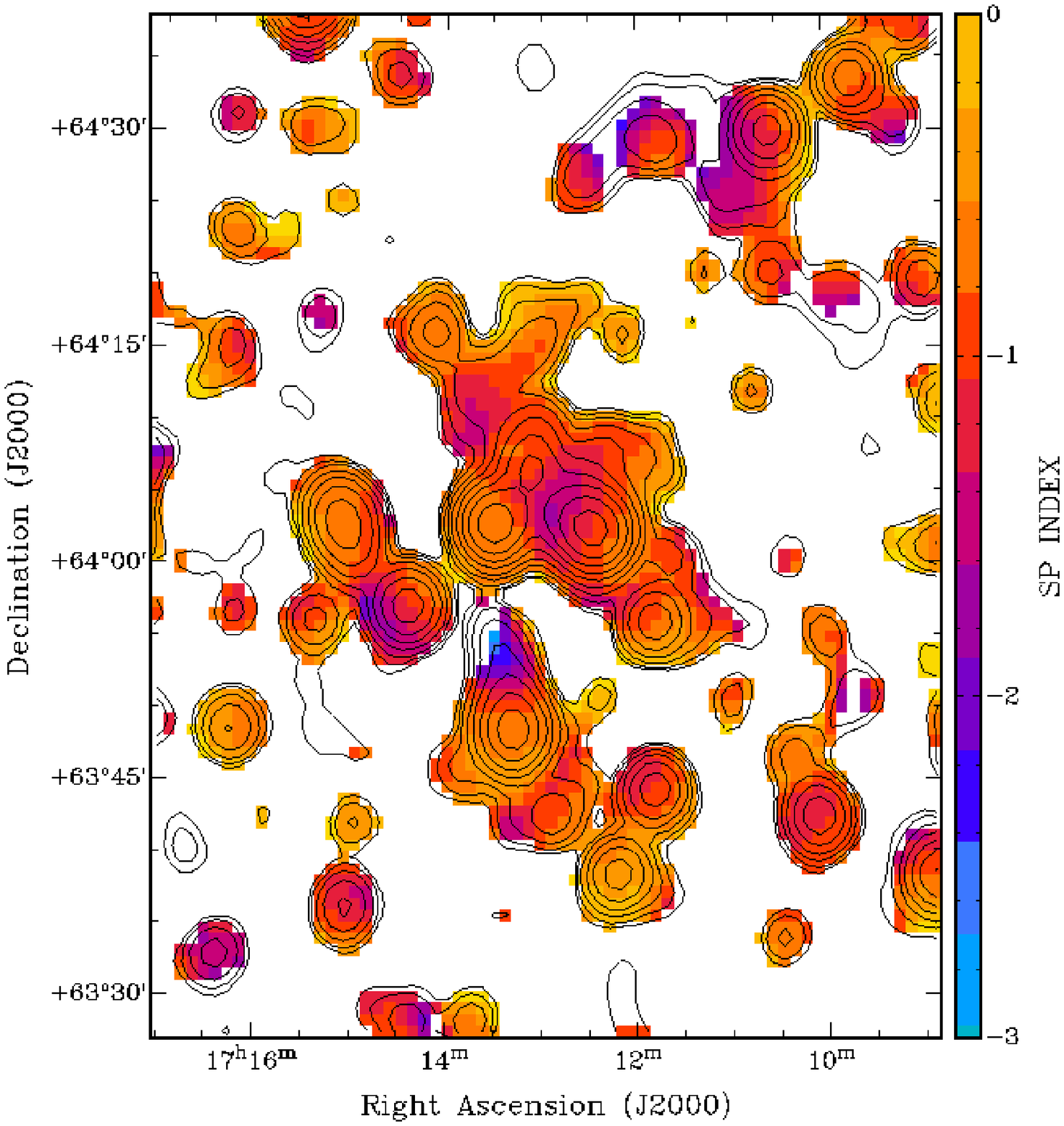}
\vspace {.4cm}
\caption{Spectral index map of A2255 between 25 cm and 85 cm, with a
resolution of $163\arcsec \times 181\arcsec$. Pixels whose brightness
was below 3$\sigma$ at 25 cm or 85 cm have been blanked. The cut is
driven by the 85 cm image in most of the points. Contour levels are
the ones of the radio map at 85 cm at low resolution: 0.0015
(3$\sigma$), 0.003, 0.006, 0.012, 0.024, 0.048, 0.096, 0.18, 0.36,
0.72, 1.4 Jy/beam.}
\label{spix2585}
\end{center}
\end{figure*}

\begin{figure*}[!]
\begin{center}
\includegraphics[scale=.63,angle=270]{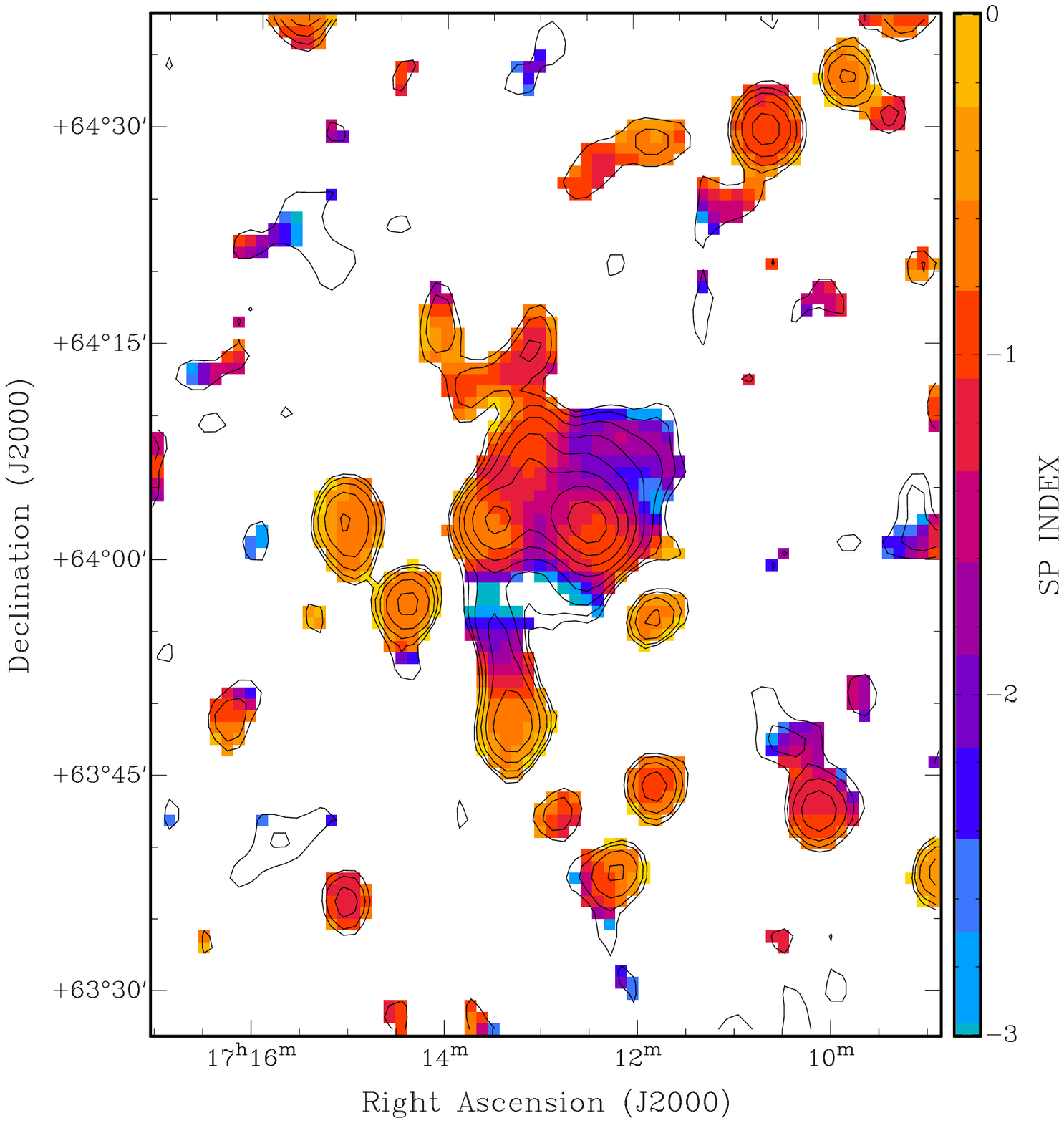}
\vspace {.4cm}
\caption{Spectral index map of A2255 between 85 cm and 2 m, with a
resolution of 163$\arcsec \times 181\arcsec$. Pixels whose brightness
was below 3$\sigma$ at 85 cm or 2 m have been blanked. The cut is
driven by the 2 m image in most of the points. Contour levels are the
ones of the radio map at 2 m: 0.015 (3$\sigma$), 0.030, 0.06, 0.12,
0.24, 0.48, 0.96, 1.8, 3.6 Jy/beam.}
\label{spix852}
\end{center}
\end{figure*}

\begin{table}
\caption{Parameters of the halo and the relics derived from the full resolution maps.}
\label{haloandrelic}
\smallskip
\begin{center}
\begin{minipage}{\textwidth}
{\small
\begin{tabular}{cccc}
\hline
\hline
\noalign{\smallskip}
Source       &         Wavelength   &    Angular size               &                       Flux                                           \\
             &                      &    ($^{\prime}$)              &                       (mJy)                                         \\     
\noalign{\smallskip}
\hline
\noalign{\smallskip}
             &            25 cm     &    7  $\times$ 4              &                      44  $\pm$ 1 $^a$                                     \\
halo         &            85 cm     &    17 $\times$ 15              &                      496 $\pm$ 7  $^a$                                    \\
             &            2  m      &   (17 $\times$ 15)$^b$        &                      805 $\pm$ 40 $^a$                                    \\
\noalign{\smallskip}
\hline
\noalign{\smallskip}
             &            25 cm     &    11 $\times$ 3              &                      30  $\pm$ 1                                      \\
NE relic     &            85 cm     &    15 $\times$ 4              &                      117 $\pm$ 2                                       \\
             &            2  m      &   (15 $\times$ 4)$^b$       &                      287 $\pm$ 17                                      \\
\noalign{\smallskip}
\hline
\noalign{\smallskip}
NW relic     &           85 cm      &    21  $\times$ 1             &                      61  $\pm$  1                                      \\ 
\noalign{\smallskip}
\hline
\noalign{\smallskip}
SW relic     &           85 cm      &    13 $\times$ 1              &                      17  $\pm$  1                                      \\  
\noalign{\smallskip}
\hline
\noalign{\smallskip}
\noalign{\smallskip}
\multicolumn{4}{l}{$^a$ The fluxes have been determined avoiding the radio galaxies}\\
\multicolumn{4}{l}{$^b$ Assumed to be the same as at 85 cm. The low resolution of }\\
\multicolumn{4}{l}{~~ the 2 m map prevented a more precise determination of this}\\  
\multicolumn{4}{l}{~~ parameter}
\end{tabular}
}
\end{minipage}
\end{center}
\end{table}

\section{Spectral index analysis of A2255}
\label{spectralanalysis}

After obtaining the final maps with the same resolution, uv min and
gridding, we computed the spectral index images using the NRAO
Astronomical Image Processing System (AIPS) package.

\subsection {Spectral index analysis between 25 cm and 85 cm}
\label{analysis2585}

The spectral index image of A2255 between 25 cm and 85 cm is shown in
Fig. \ref{spix2585}.\\
We did not subtract the extended discrete radio sources, to avoid
errors which might be introduced by the residuals of the
subtraction. Therefore, we notice that the steep $\alpha$ values in
the central area of the halo are probably associated with the tail of
the ``Original TRG''. Moreover, the flat spectrum of the eastern
regions of the halo, are heavily influenced by the presence of the
nuclei of the Double and the Goldfish radio galaxies. For this reason,
the only $\alpha$ values in the N and NW area of the halo can be
considered as representative of the halo itself.  Here, the spectrum
shows a radial flattening from the center ( $\alpha \sim -1.6 \pm
0.05$) to the periphery ($\alpha \sim -0.7 \pm 0.07$), where the three
bright radio filaments are detected at high-frequency (see
Fig. \ref{25cmfig}). At this location, fresh (re-)acceleration of
relativistic particles is expected to take place, producing a
flattening of the spectrum.

\begin{figure*}
\begin{center}
\includegraphics[scale=.57,angle=0]{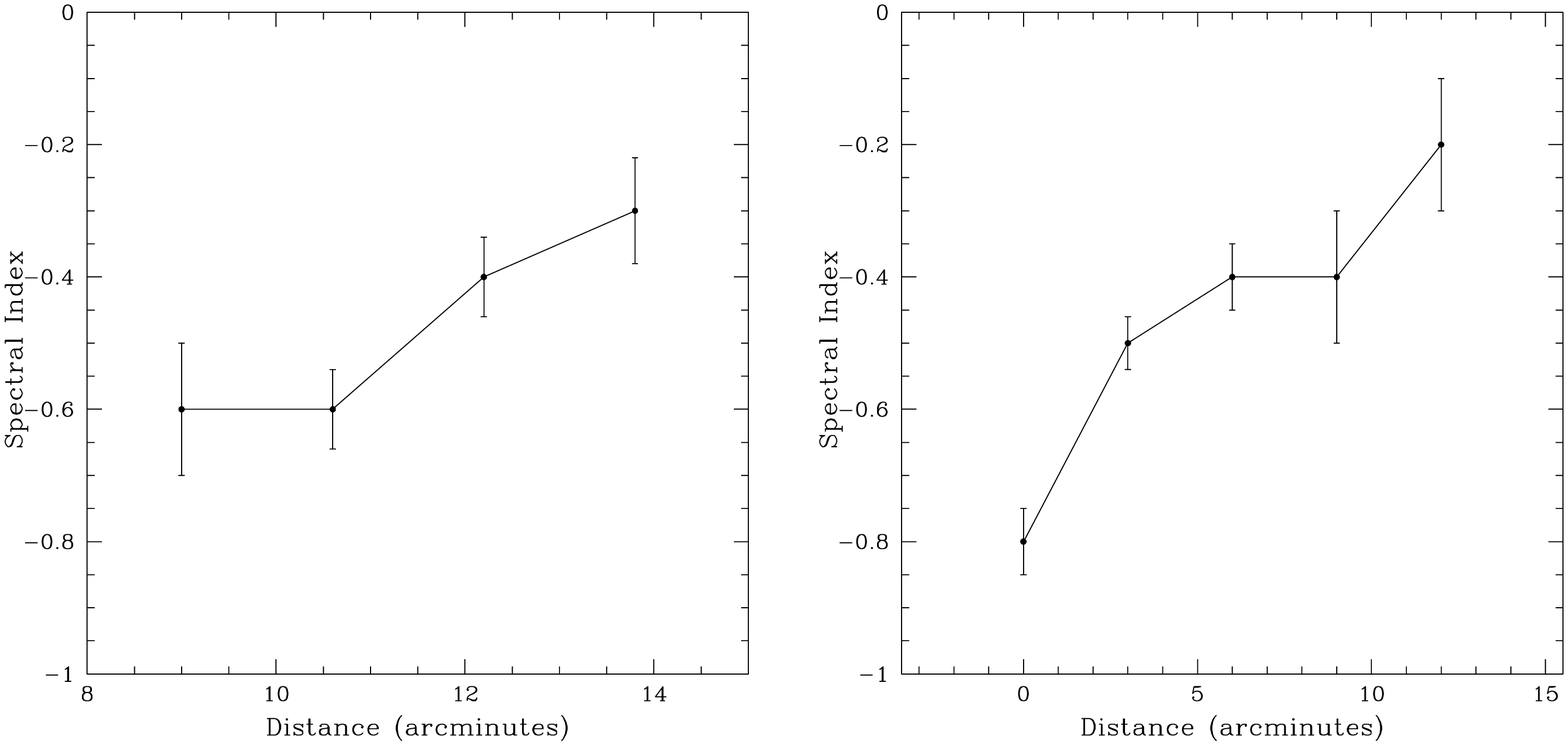}
\vspace {.4cm}
\caption{Spectral index profile of the NE relic between 25 cm and 85
cm along the minor axis (left panel) and the main axis (right
panel). For the former the distance is from the cluster center, for
the latter the distance is from position ${\rm RA} = 17^{\rm h}13^{\rm m}
51^{\rm s}$, ${\rm Dec} = +64\hbox{$^{\circ}$}12\hbox{$^{\prime}$}00\hbox{$\arcsec$}$, which is
the SE edge of the relic.}
\label{spixprofileNERELIC}
\end{center}
\end{figure*}

The NE relic shows $\alpha$ values which range from $-1.46 \pm 0.08$
to $-0.4 \pm 0.09$. There is a gradient of the spectral index along
the main axis: $\alpha$ flattens from south-east towards north-west
(see right panel of Fig. \ref{spixprofileNERELIC}). Moreover, $\alpha$
shows a trend also along the minor axis, being steeper in the regions
close to the cluster center and flattening towards the periphery (see
left panel of Fig. \ref{spixprofileNERELIC})

As discussed in Sect. \ref{25cmmap}, the brighter filament of the NW
relic is detected at 25 cm as a very low brightness feature
(1$\sigma$/beam). Because of the 3$\sigma$ criterium we used to
produce the spectral index map between 25 cm and 85 cm, this feature
is not visible in the low resolution map.

\begin{figure*}
\begin{center}
\includegraphics[scale=.4,angle=0]{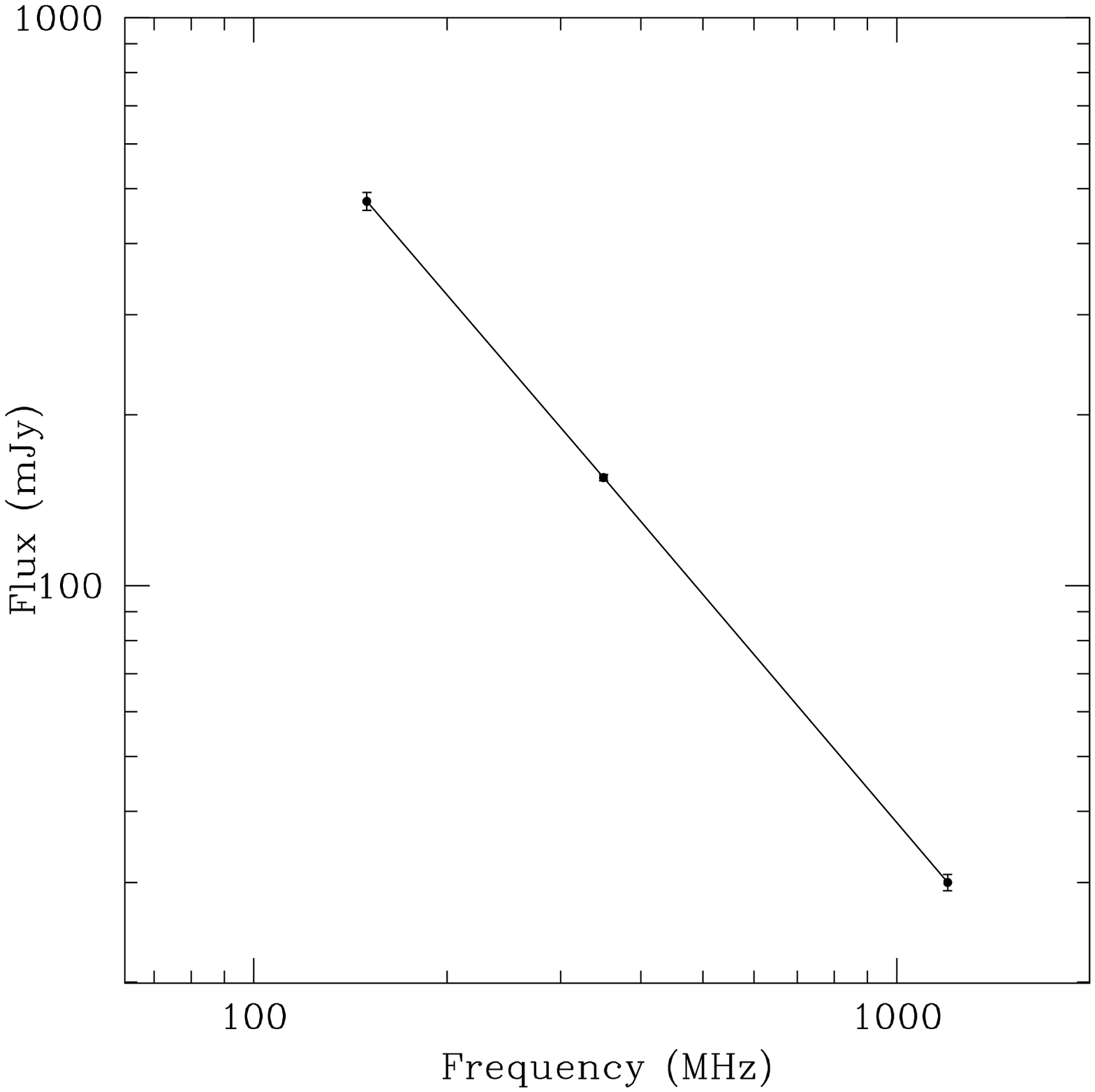}
\includegraphics[scale=.4,angle=0]{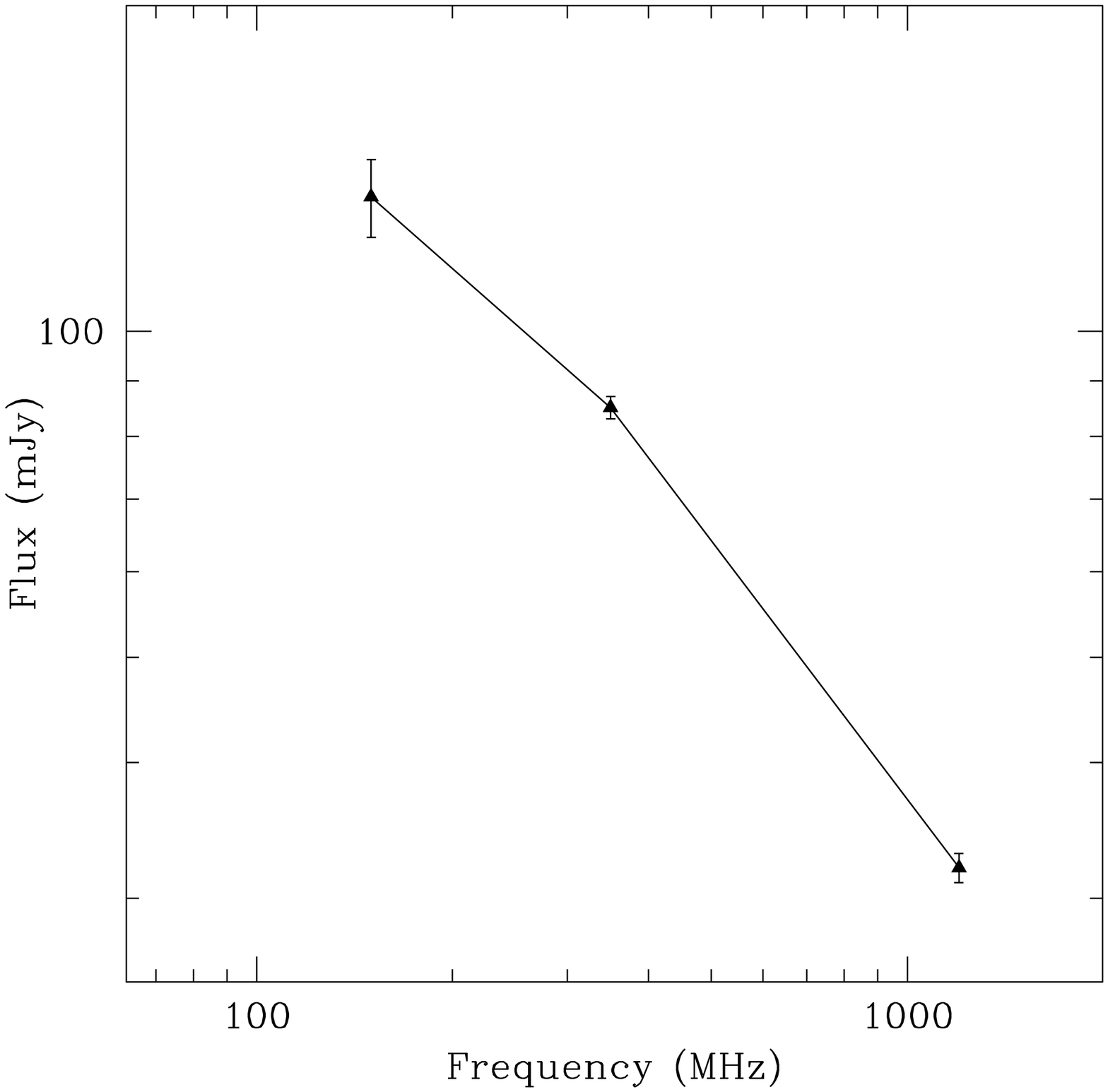}
\caption{Radio spectra of the integrated emission of the radio halo
(left panel) and of the NE radio relic (right panel) for the frequency
range 2 m--25 cm. For this spectral comparison, we have computed the
fluxes of the halo in its NW region.}
\label{integratedspectrahalo&relic}
\end{center}
\end{figure*}

\subsection {Spectral index analysis between 85 cm and 2 m}
\label{analysis852}

Fig. \ref{spix852} shows the spectral index image of A2255, calculated
between 2 m and 85 cm.  The spectral index image has been obtained
clipping the pixels where the brightness was below $3\sigma$ level at
both wavelengths.

The spectral index within the halo shows a more patchy structure than
between 25 cm and 85 cm. The $\alpha$ values are flatter in the regions corresponding to
the nuclei of the central radio galaxies ($\alpha = -0.9 \pm 0.01$)
and steepen in between ($\alpha = -1.6 \pm 0.01$). In the NW
regions of the halo, where there is no contamination from
radio galaxies, $\alpha$ shows a discontinuous behavior, ranging
between $-2.8 \pm  0.1$ and  $-1.6 \pm 0.07$. As for
the spectral index image between 25 cm and 85 cm, the flatter values
are confined to a semi circular area that approximately describes the
location of the 3 bright filaments at high-frequency.

The NE relic shows $\alpha$ values in the range of $-0.9 \pm 0.1$ to
$-1.2 \pm 0.1$, with the steeper values confined in its NW areas. We
notice, however, that because of the criteria used to make the
spectral index map, combined with the high noise level in the central
part of the 2 m map, $\alpha$ has been determined for the central area
of the NE relic only. Moreover, given the high rms values associated
with the spectral index in its SE region, no clear radial trend
neither along the main axis nor perpendicular to it can be determined.

The NW relic, detected at 2 m in the high sensitivity map, is no
longer entirely visible and only a small part of the filament pointing
towards the cluster center is still detected. Its spectral index
ranges between --1.2$\pm$ 0.2 and --0.5$\pm$ 0.1, and steepens towards
the cluster center. Moreover, there is a feature at location ${\rm RA} =
17^{\rm h}11^{\rm m} 00^{\rm s}$, ${\rm Dec} = +64\hbox{$^{\circ}$
}24\hbox{$^{\prime}$}49\hbox{$\arcsec$}$, which is likely associated
with the second filament perpendicular to the previous one. Here,
$\alpha$ steepens towards the cluster center and ranges between $-1.8
\pm 0.2$ and $-0.6 \pm 0.3$.

We determined the integrated synchrotron spectrum of the region of the
radio halo without the contamination of radio galaxies and of the NE
relic in the wavelength range 2 m--25 cm
(Fig. \ref{integratedspectrahalo&relic}). The fluxes of the analyzed
features are reported in Table \ref{halo&relic}. The spectrum of the
radio halo has a constant slope between the 3 wavelengths ($\alpha_{85
cm}^{2 m} = -1.30 \pm 0.05$ and $\alpha_{85 cm}^{25 cm} = -1.30 \pm
0.1$), while the one of the NE relic is flatter at low-frequency
($\alpha_{85 cm}^{2 m} = -0.5 \pm 0.2$) and steepens at high-frequency
($\alpha_{85 cm}^{25 cm} = -0.81 \pm 0.1$).

\begin{table*}
\caption{Parameters of the NW region of
the halo and of the NE relic derived from the images produced with the
same uv coverage and the same resolution.}
\label{halo&relic}
\smallskip
\begin{center}
{\small
\begin{tabular}{ccccccccccc}
\hline
\hline
\noalign{\smallskip}
Region Flux & RA$_{J2000}$ &  Dec$_{J2000}$                         & Size     &$S_{tot} ^{\lambda = 2 m}$ &  $S_{tot} ^{\lambda = 85 cm}$  & $S_{tot} ^{\lambda = 25 cm}$    & $\alpha_{85cm} ^{2 m}$  & $\alpha_{25 cm} ^{85 cm}$ & $u_{min}$   & $B_{eq}$   \\
            & h,m,s        & $^{\circ}$, $^{\prime}$, $\arcsec$     & (kpc)    &  mJy                      &        mJy                     &        mJy                      &                         &                           &(erg/$cm^3$) & ($\mu$G)  \\
\noalign{\smallskip}
\hline
\noalign{\smallskip}
Radio Halo $^a$  & 17 13 00     & 64 07 59                               & 630$\times$360 & 475 $\pm$ 17            &  155 $\pm$ 2                  &      30 $\pm$ 1                &   $-1.3 \pm 0.05$           & $-1.3 \pm 0.1$                         &3.3 10$^{-14}$& 0.6  \\
\noalign{\smallskip}
\hline
\noalign{\smallskip}
NE Radio Relic  & 17 13 22 & 64 13 36                               & 990$\times$270 & 133 $\pm$  11           &  85  $\pm$ 1                  &      32   $\pm$ 1              &   $-0.5 \pm 0.1$            &  $-0.8 \pm 0.1$                        &1.4 10$^{-15}$ & 0.4  \\ 
\noalign{\smallskip}
\hline
\noalign{\smallskip}
\noalign{\smallskip}
\multicolumn{11}{l}{$^a$ For the halo the parameters have been determined
  avoiding the radio galaxies}
\end{tabular}
}
\end{center}
\end{table*}

\subsection {Equipartition magnetic field}
\label{equipartitionmagneticfield}

The total energy content of a synchrotron source ($U_{tot}$) is given
by the contribution of the energy of the relativistic particles and
the energy of the magnetic field. $U_{tot}$ shows a minimum when these
2 contributions are approximately equal. For this reason the minimum
energy is known also as equipartition value. If we assume that a radio
source is in a condition of minimum energy, it is possible to estimate
the magnetic field strength \citep{1970ranp.book.....P}. The
equipartition magnetic field ($B_{eq}$) is expressed by
\begin{equation}
B_{eq}=\left( \frac{24\pi}{7} u_{min}\right )^{1/2} \; ,
\end{equation} 
where the total minimum energy density ($u_{min}$) of the source
depends on observational quantities, as the source brightness (I$_0$),
its redshift (z), and the observing frequency ($\nu_0$), and on unknown
parameters, like the ratio between the energy in relativistic protons
and electrons (k) and the filling factor ($\Phi$), which represent the
fraction of the source volume that is filled with particles and
magnetic fields :
\begin{eqnarray}
\label{equip}
u_{min} \left[ \frac{{\rm erg}}{{\rm cm^3}} \right ] = \xi(\alpha,\nu_1,\nu_2) (1+k)^{4/7} (\nu_{0[{\rm MHz}]})^{-4\alpha/7} (1+z)^{(12 - 4\alpha)/7} \times \nonumber    \\
\times \left( I_{0\left[\frac{{\rm mJy}}{{\rm arcsec^2}}\right]}\right)^{4/7} (d_{[{\rm kpc}]})^{-4/7} \Phi^{-4/7} \;\;\;\;\;\;\;\;\;\;\;\;\;\;\;\;\;\;\;\;\;\;\;\;\;\;\;
\end{eqnarray}
Here, $\xi(\alpha,\nu_1,\nu_2)$ is a constant which is tabulated in
\citet{2004IJMPD..13.1549G} and $d$ is the source depth. The latter formula is mainly used for significantly extended structures. In the case of compact features, equation \ref{equip} is better expressed by
\begin{eqnarray}
u_{min} \left[ \frac{{\rm erg}}{{\rm cm^3}} \right ] = \xi(\alpha,\nu_1,\nu_2) (1+k)^{4/7} (\nu_{0[{\rm MHz}]})^{-4\alpha/7} (1+z)^{(12-4\alpha)/7} \times \nonumber        \\
 \times {\left(I_{0[{\rm mJy}]}\right)}^{4/7} {\left({\rm V}_{[{kpc}
       ^3]}\right)}^{-4/7} \Phi^{-4/7} \;\; , \;\;\;\;\;\;\;\;\;\;\;\;\;\;\;\;\;\;\;\;\;\;\;\;\;\;
\end{eqnarray}
where V is the source volume.\\ Table \ref{halo&relic} lists the equipartition
magnetic field for the region of the halo free of radio galaxies and for the
NE relic. It was computed using the
fluxes at 85 cm. Furthermore, we assumed equal energy in relativistic protons
and electrons ($k$ = 1) and a filling factor of unity ($\Phi$ = 1, the volume is
homogeneously filled by the relativistic plasma). The synchrotron luminosity
is calculated in the frequency range 10 MHz--100 GHz.  Halo and relic are
shaped as cylinders with a size (length $\times$ depth, derived from the map at 25 cm) of $630 \times 360$
kpc and $990 \times 270$ kpc, respectively. A spectral index of $\alpha=-1.3$ for the halo
and $\alpha=-0.7$ for the NE relic was adopted.

\citet{2006A&A...460..425G} studied the intra cluster magnetic field
power spectrum of A2255 through the analysis of the RM distributions of three
cluster radiogalaxies. They found that to reproduce
the behavior of the RM for the radio galaxies, the power spectrum of
the magnetic field of the cluster should steep from the center to the
periphery, with an average magnetic field strength for the radio halo
calculated over 1 Mpc$^3$ of about 1.2 $\mu$G. This value is a factor
of 2 higher than the equipartition magnetic field obtained by us. The
discrepancy is mainly due to the fact that, in the approach followed
by \citet{2006A&A...460..425G}, also low energy relativistic particles
are taken into account when computing the strength of the magnetic
field, while in the standard approach used in the computation of the
equipartition parameters one considers a cut frequency window between
$\nu_{1}$ and $\nu_{2}$. \citet{1997A&A...325..898B} demonstrated that
it is more appropriate to calculate the radio source energy by
integrating the synchrotron luminosity over a range of electron
energies. This method has the advantage that electrons of very low
energy are also taken into account and it avoids the problem that
electron energies corresponding to frequencies $\nu_1$ and $\nu_2$
depend on the magnetic field value. Representing the electron energy
with its Lorentz factor $\gamma$, and assuming that $\gamma_{min}$
$\ll$ $\gamma_{max}$, the new expression of the equipartition magnetic
field is
\begin{eqnarray}
\label{brunetti}
B^\prime_{eq} \sim 1.1 \gamma_{min}^{\frac{1+2\alpha}{3-\alpha}} B_{eq}^{\frac{7}{2(3-\alpha)}}\;\;\;, 
\end{eqnarray}
where $B^\prime_{eq}$ and $B_{eq}$ are expressed in Gauss.  Using
equation \ref{brunetti}, we find that the equipartition magnetic field
computed by us and the average magnetic field of the halo calculated
by \citet{2006A&A...460..425G} are in agreement if we assume a cut in
the energy distribution of the relativistic particles at $\gamma_{min}
= 310$. By using the same $\gamma_{min}$ as for the radio halo, the
new equipartition magnetic field for the NE relic is $\sim$ 0.6
$\mu$G.

\section{The head-tail Beaver radio galaxy}
\label{beaver}

Abell 2255 is one of the richest clusters of the Abell catalog in
terms of radio galaxies. In our maps we can detect 7 radio galaxies:
four are located in the central region of the cluster, and three lie
at large projected distance from it. The physical properties of these
radio galaxies have already been studied by several authors
\citep{har,fer,2003AJ....125.2427M}, but so far spectral index images
were not presented in the literature.  During the analysis of our
multi-frequency observations, we studied the physical properties of
the Beaver radio galaxy because its long tail gave us the possibility
to test ageing models of the radiating electrons.

\begin{figure*}
\begin{center}
\includegraphics[scale=.80,angle=0]{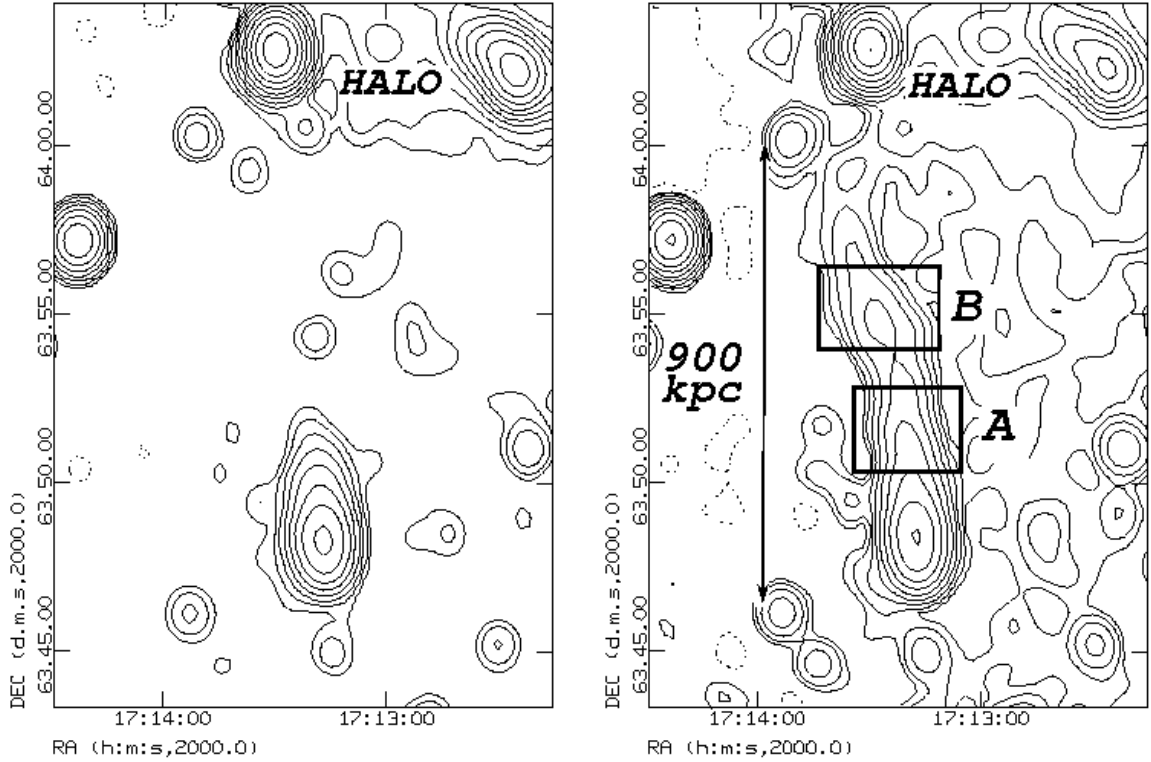}
\caption{Comparison between the morphologies of the Beaver radio
galaxy at 25 cm (left panel) and 85 cm (right panel). The 25 cm map
has been convolved with the 85 cm beam (54$\arcsec \times
64\arcsec$). Boxes A and B define two regions along the tail where we
did our analysis (see
Sect. \ref{physicalparametersbeaver}--\ref{spectralageing}). At 25 cm,
the contours are --0.25 , 0.25,0.50, 1, 2, 4, 8, 16, 32, 64, 128
mJy/beam; at 85 cm, --0.5, 0.5, 1, 2, 4, 8, 16, 32, 64, 128, 256, 512
mJy/beam. The images are not corrected for the primary beam.}
\label{comparisonbeaver}
\end{center}
\end{figure*}

\begin{figure}[t!]
\begin{center}
\includegraphics[scale=.50,angle=270]{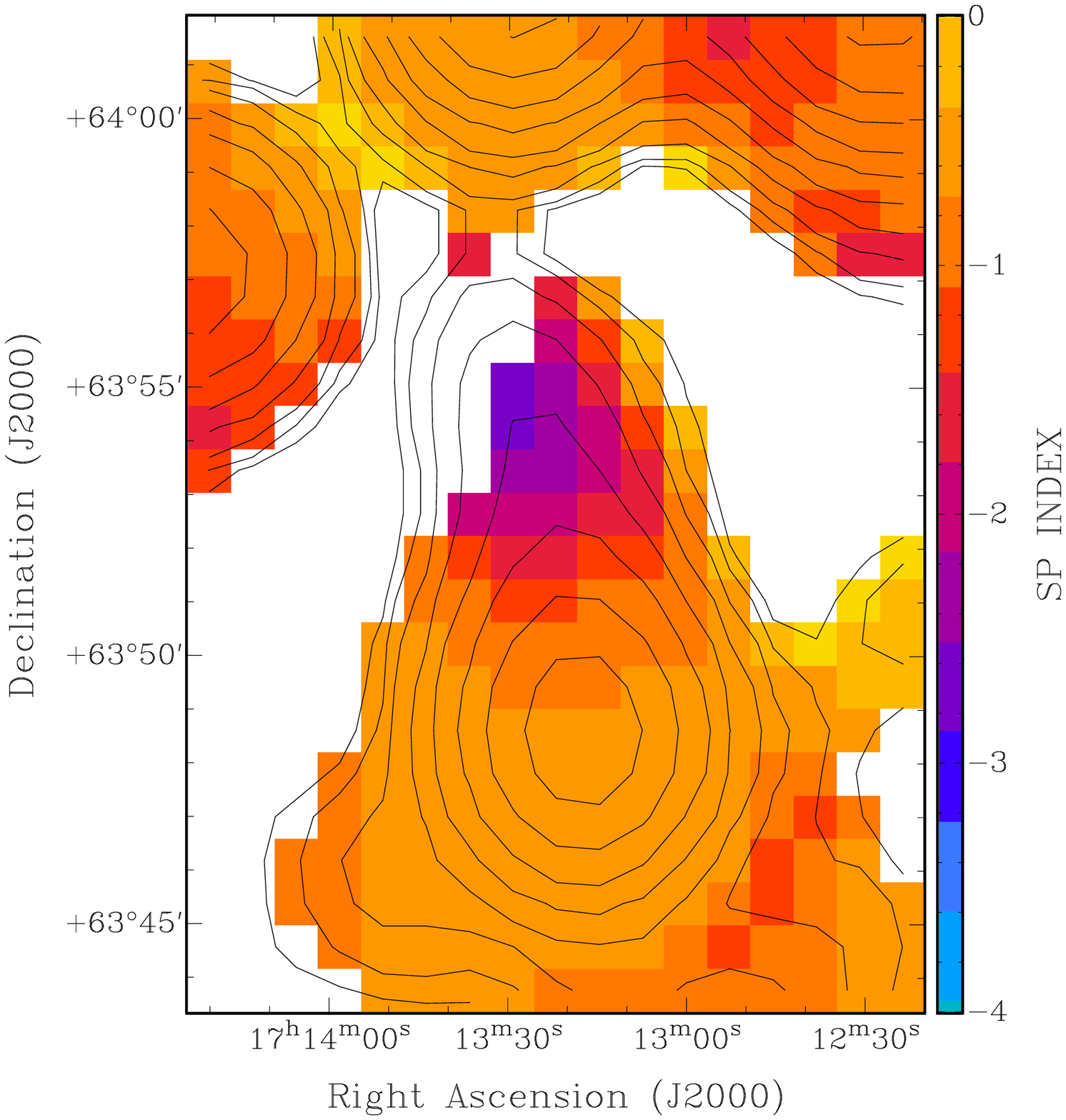}
\vspace{0.08cm}
\caption{Spectral index map of the Beaver radio galaxy between 25 cm
and 85 cm, with a resolution of 163$\arcsec \times 181
\arcsec$. Pixels whose brightness was below 3$\sigma$ at 25 cm or 85
cm have been blanked. The cut is driven by the 85 cm image in most of
the points. Contour levels are the ones of the radio map at 85 cm at
low resolution: 0.0015 (3$\sigma$), 0.003, 0.006, 0.012, 0.024, 0.048,
0.096, 0.18, 0.36, 0.72, 1.4 Jy/beam.}
\label{beaverspix2585}
\end{center}
\end{figure}
\vspace{2cm}

\begin{figure}
\begin{center}
\includegraphics[scale=.50,angle=270]{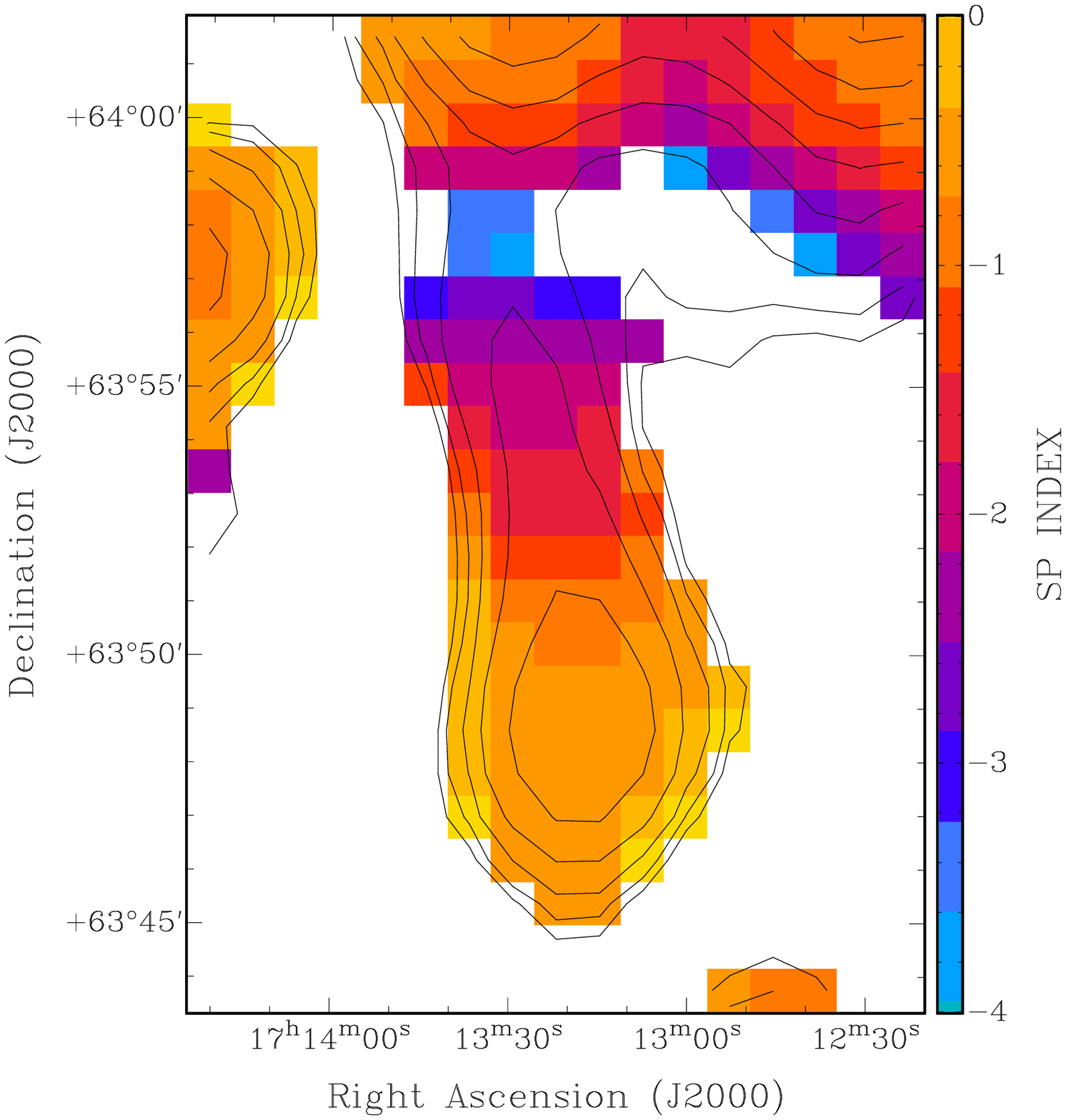}
\vspace{0.08cm}
\caption{Spectral index map of the Beaver radio galaxy between 85 cm
and 2 m, with a resolution of 163$\arcsec \times 181 \arcsec$. Pixels
whose brightness was below 3$\sigma$ at 85 cm or 2 m have been
blanked. The cut is driven by the 2 m image in most of the
points. Contour levels are the ones of the radio map at 2 m: 0.015
(3$\sigma$), 0.030, 0.06, 0.12, 0.24, 0.48, 0.96, 1.8, 3.6 Jy/beam.}
\label{beaverspix852}
\end{center}
\end{figure}
\subsection{Spectral index analysis}
\label{spectralanalysisbeaver}

One of the most interesting properties of the Beaver is that it
doubles the length of its tail to almost 1 Mpc between 25 cm and 85 cm
(see Fig. \ref{comparisonbeaver}). This suggests very steep $\alpha$
values for the ending part of the tail, which is an indication of
severe energy losses suffered by the relativistic particles.

\begin{figure*}
\begin{center}
\includegraphics[scale=.40,angle=0]{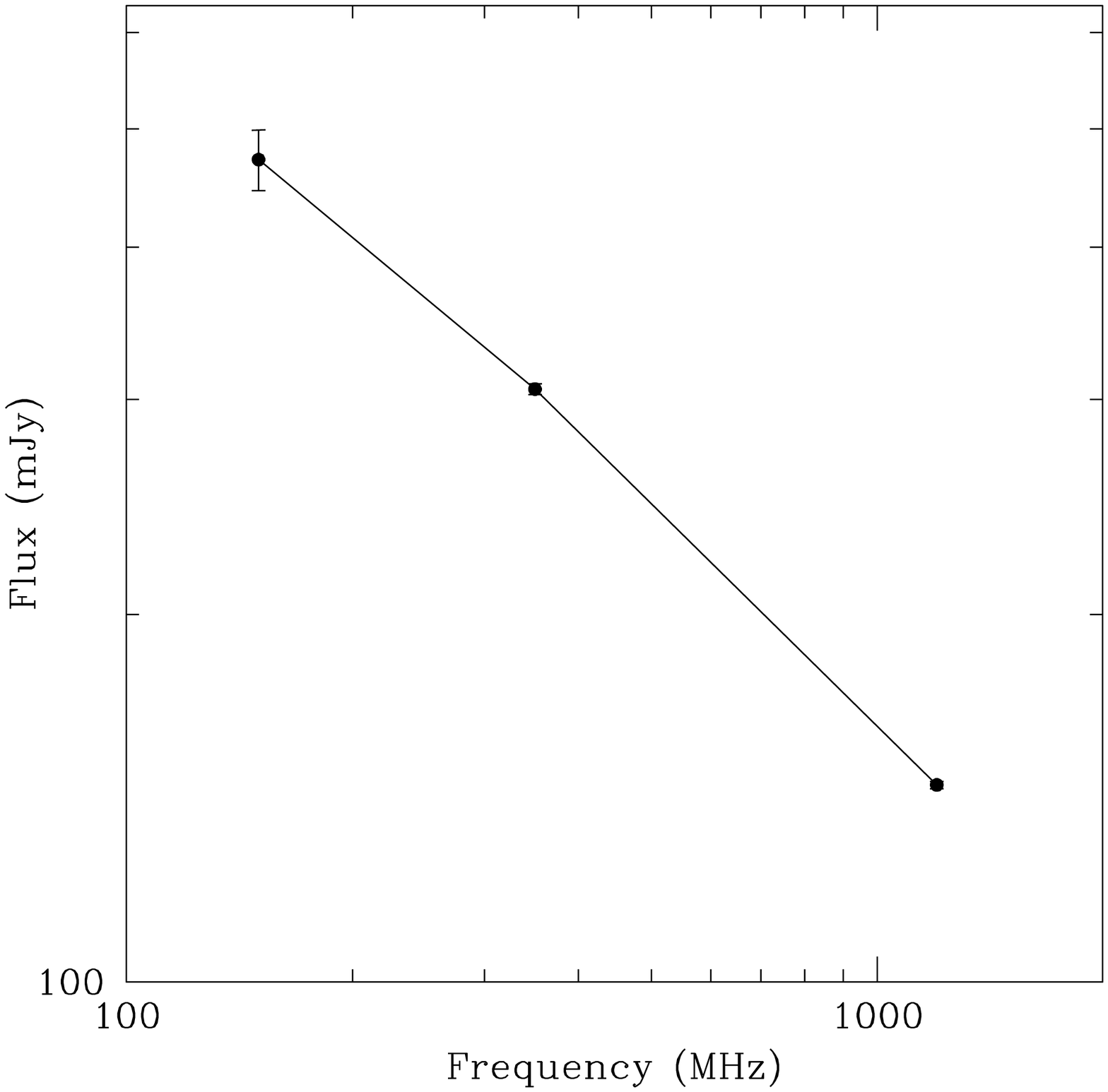}
\includegraphics[scale=.40,angle=0]{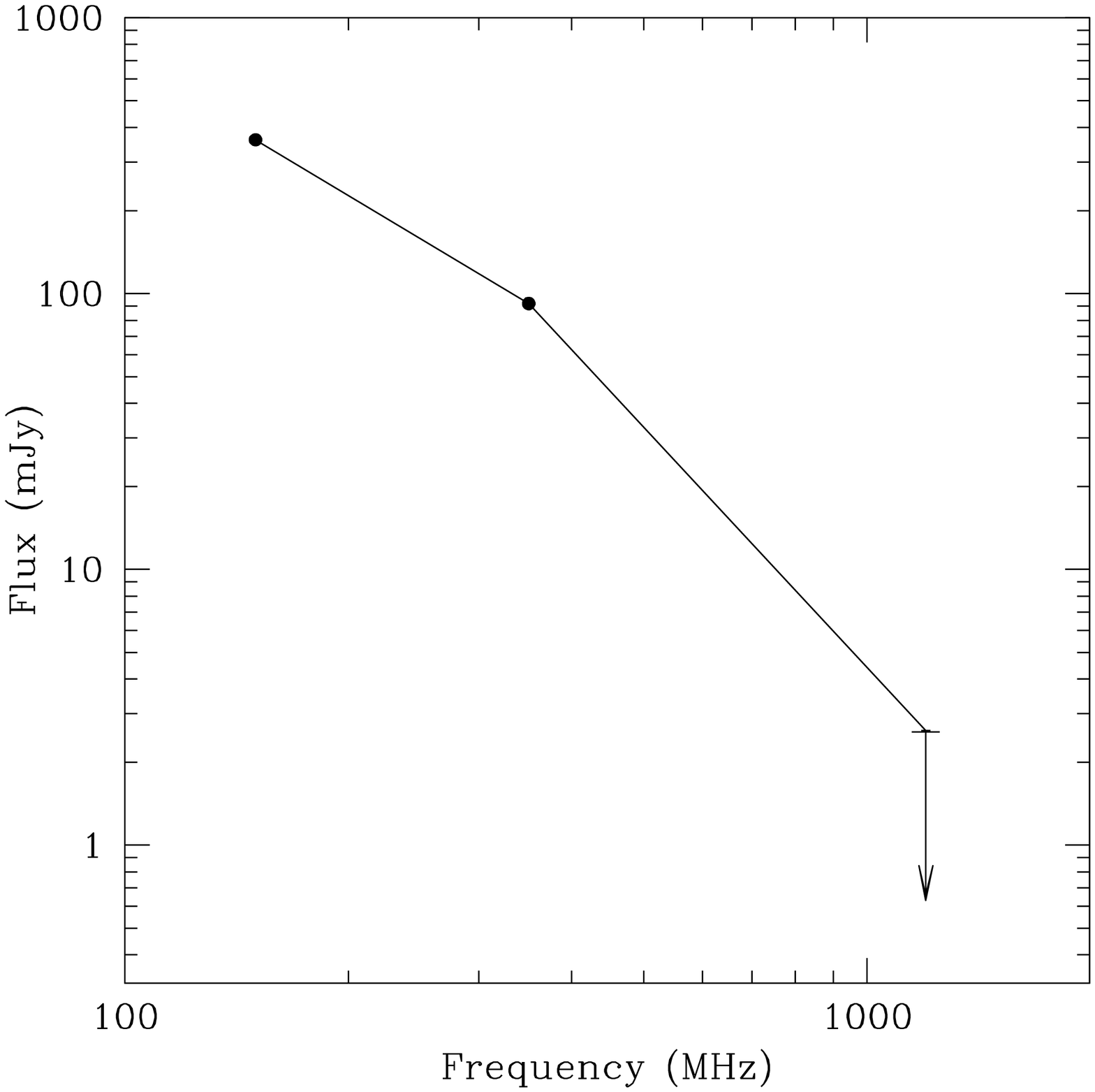}
\caption{Radio spectra of the head (left panel) and the tail (right
panel) of the Beaver radio galaxy in the frequency range 2 m--25 cm.}
\label{integratedspectrabeaver}
\end{center}
\end{figure*}

\begin{figure*}
\begin{center}
\includegraphics[scale=.40,angle=0]{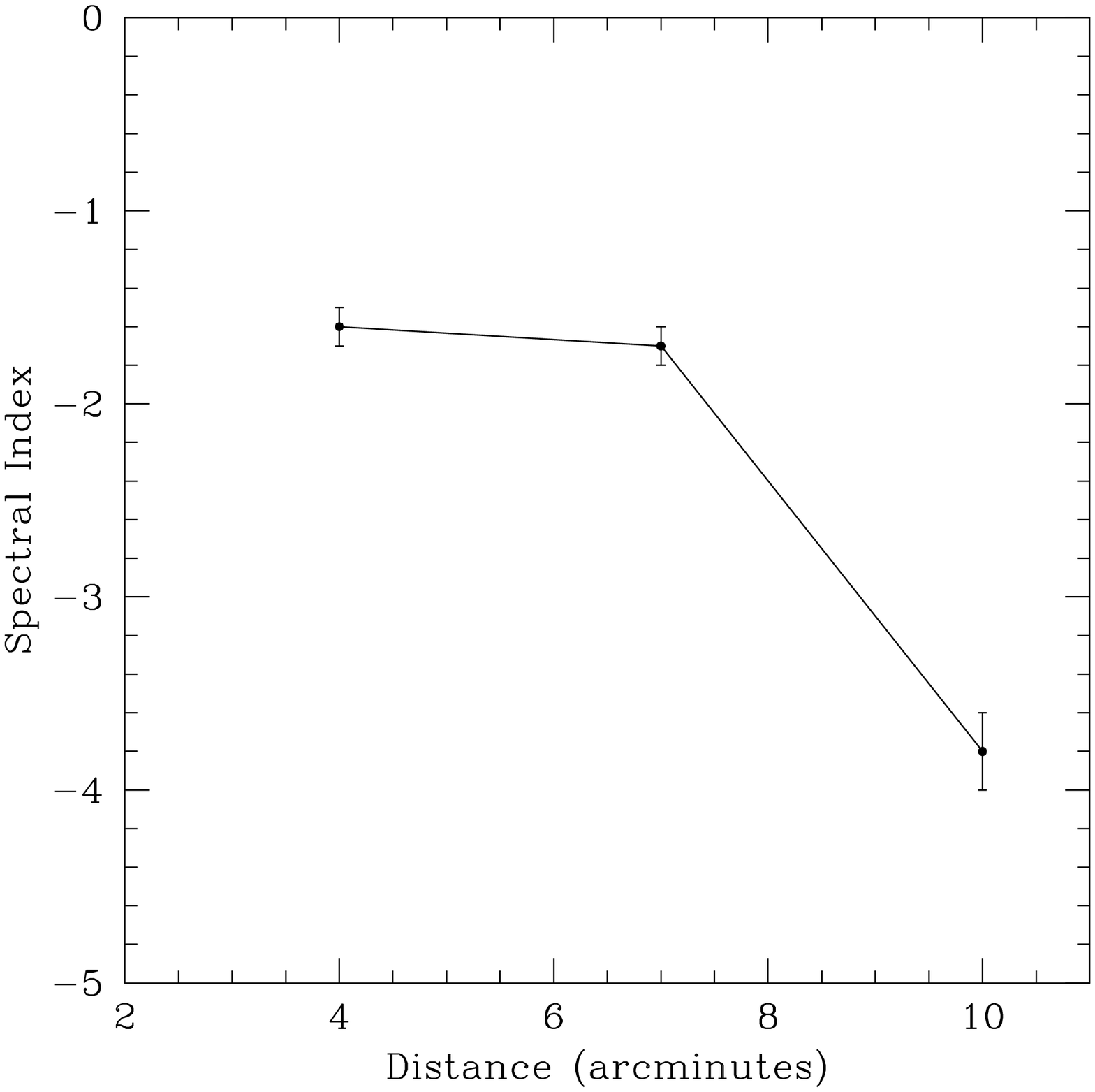}
\includegraphics[scale=.40,angle=0]{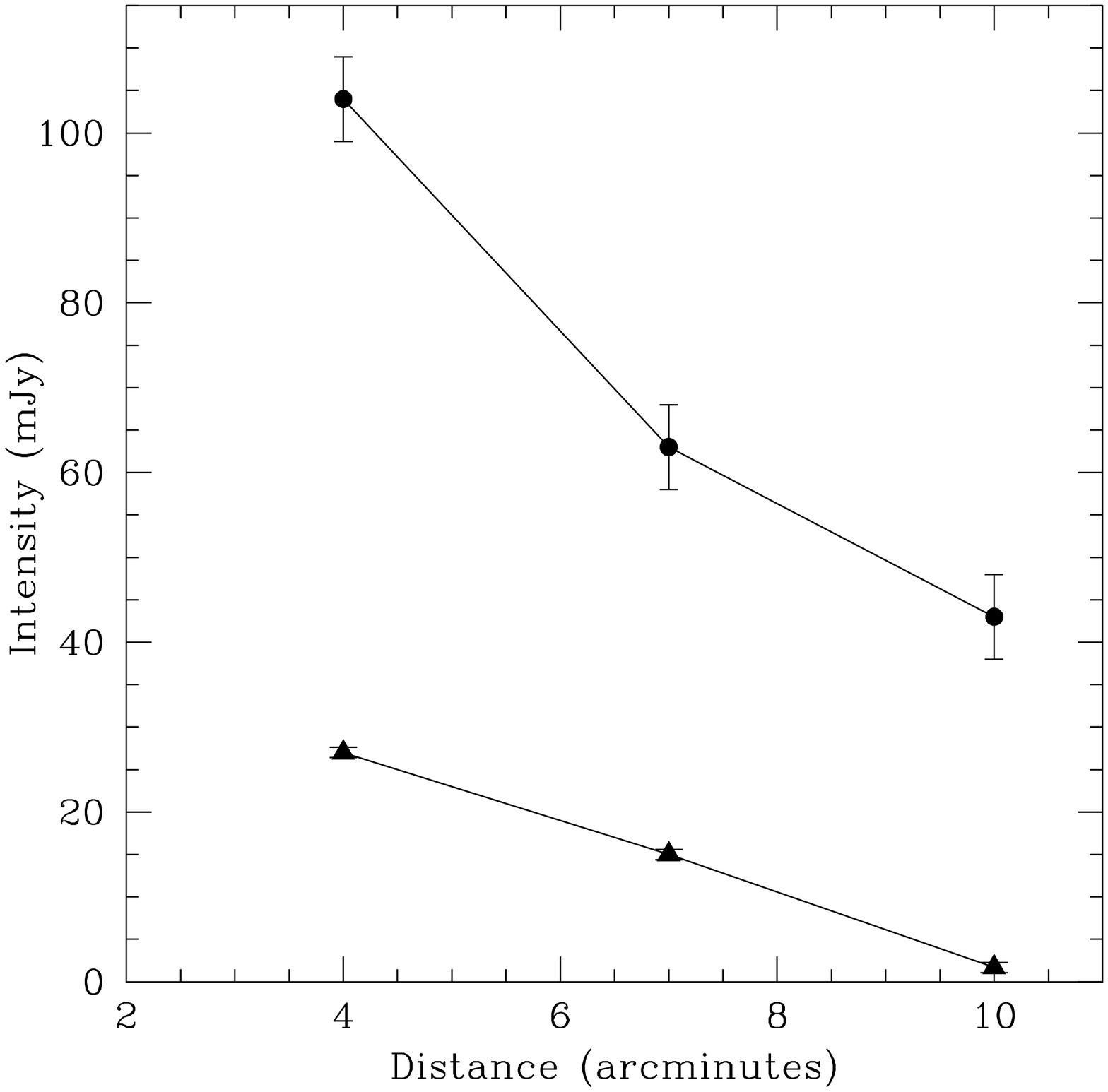}
\caption{Right panel:point to point spectral index between 85 cm and 2
m. The $\alpha$ values are plotted versus the distance from the
nucleus of the Beaver, in arcminutes. Right panel: Variation of the
primary beam corrected brightness at 2 m (top line) and 85 cm (bottom
line).}
\label{beaverprofiles}
\end{center}
\end{figure*}

In Figs. \ref{beaverspix2585} and \ref{beaverspix852}, we show the
spectral index maps of the Beaver, between 25 cm and 85 cm, and
between 85 cm and 2 m. The spectral index distribution clearly
confirms that we are dealing with a tailed radio galaxy, since
$\alpha$ severely steepens going from the head of the source towards
the end of the tail. Between 25 cm and 85 cm, the spectral index of
the Beaver could be computed for the head and the initial part of the
tail. As we expect, $\alpha$ is flatter in the regions closer to the
nucleus ($\alpha = -0.5 \pm 0.08$), where more energetic electrons are
continuously injected by the central AGN, and then it steepens
reaching values of $\alpha = -2.2 \pm 0.2$.
We would have expected a flatter spectral index for the head of the
radio galaxy, but the low resolution of the maps makes the final
spectral index of the nucleus contaminated by the surrounding steeper
regions.  The spectral index between 85 cm and 2 m shows basically the
same trend, being flatter in the nucleus region ($\alpha \sim -0.6 \pm
0.03$) and steepening towards the cluster radio halo ($\alpha \sim
-3.3 \pm 0.2$).\\ The integrated spectra of the head and the tail of
the Beaver are shown in Fig. \ref{integratedspectrabeaver}.  The trend
of the spectral index and of the primary beam corrected brightness
between 2 m and 85 cm along the tail is shown in
Fig. \ref{beaverprofiles}. We computed $\alpha$ at 3 different
positions, starting from 4$^{\prime}$ from the nucleus of the radio galaxy
and going towards the end of the tail, averaging the $\alpha$ values
within beam size boxes.

\subsection{Physical parameters and spectral ageing in the tail of the Beaver}
\label{spectralageing}

The electrons in the tail of the radio galaxies are not thought to be
subjected to large bulk motions. Their number in a particular region
of the tail remains constant after the galaxy has passed. Therefore,
their position can be considered a measure of the age of the
electrons.

Given a homogeneous and isotropic population of electrons with a power
law energy distribution ( $N(E)dE = N_{0} E^{-\gamma} dE$ ), the
synchrotron spectrum for regions optically thin to their own radiation
varies with frequency as
\begin{equation}
J(\nu) \propto N_0 (B sin \theta)^{1-\alpha} \nu^{\alpha} \; ,
\end{equation} 
where $\theta$ is the pitch angle between the electron velocity and
the magnetic field direction. The spectral index $\alpha$ is related
to the index of the electron energy distribution: $\alpha$ =
(1-$\gamma$)/2.
 
The electrons energy decreases with time and the resulting
synchrotron spectrum undergoes to a modification. One will observe a
critical frequency $\nu^{*}$, such that for $\nu$ $<$ $\nu^{*}$ the
spectrum is unchanged, whereas for $\nu$ $>$ $\nu^{*}$ the spectrum
steepens.\\

There are several mechanisms that can make the electrons loose their
energy: synchrotron radiation, inverse Compton scattering, adiabatic
expansion, Bremstrahlung and ionization losses
\citep{1962SvA.....6..317K,1970ranp.book.....P}. However, for confined
tails the first two mechanisms play the most important role.

\begin{figure*}
\begin{center}
\includegraphics[scale=.9,angle=0]{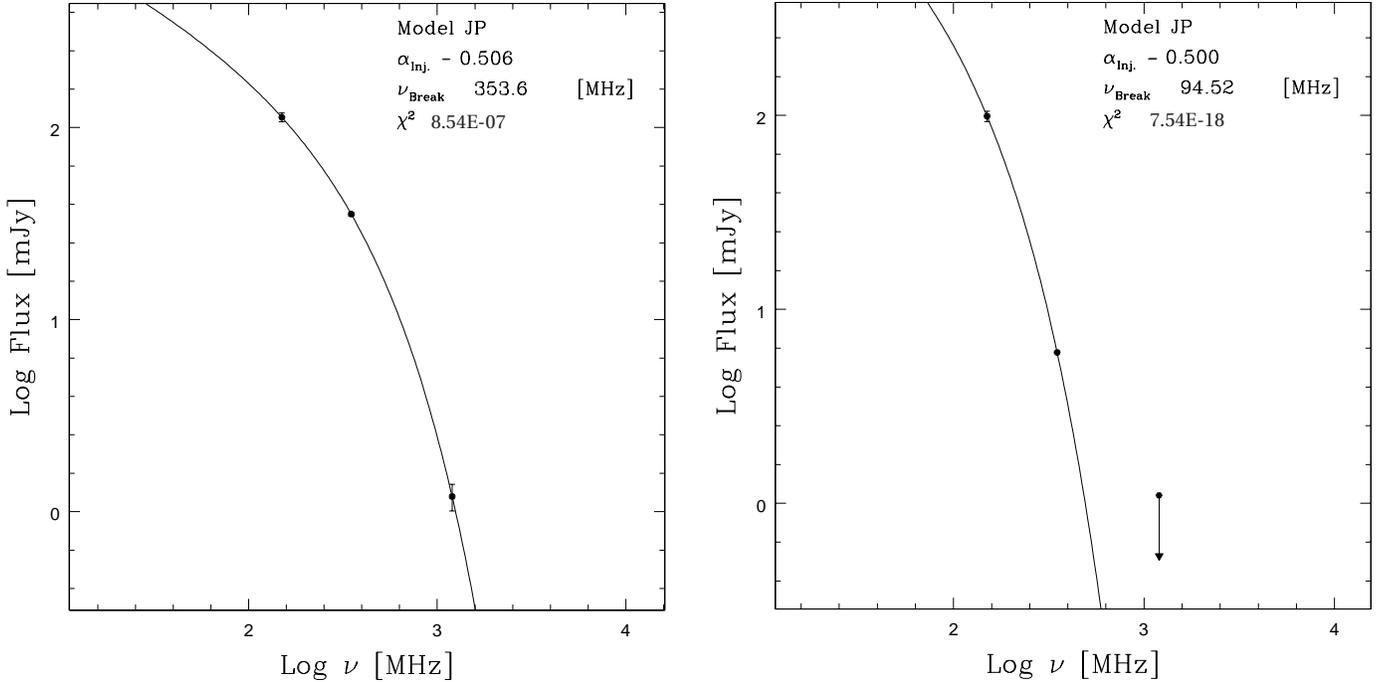}
\caption{Fluxes of region A (left panel) and region B (right panel)
  fitted with the JP model. $\chi^{2}$ represents the reduced
  $\chi^2$.}
\label{jpmodels}
\end{center}
\end{figure*}

There are mainly two models that describe the steepening of the radio
spectrum above $\nu^{*}$:
\begin{itemize}
\item {\it the Kardashev-Pacholczyk model \citep[KP model,
][]{1962SvA.....6..317K, 1970ranp.book.....P}}, in which the electrons
maintain the same pitch angle with respect to the B lines. The
particles will have different energy losses depending on the value of
this parameter;
\item {\it the Jaffe-Perola model \citep[JP model,
][]{1973A&A....26..423J}}, where one considers an electrons population
with an isotropic distributions of pitch angles. This results in a
sharp energy cutoff in the energy electron distribution, with a
synchrotron spectrum showing an exponential drop at high-frequency.
\end{itemize}
Since the electrons within a radio source are likely to have various
  distributions of the pitch angle, the JP model better describes this
  physical situation.
Ignoring adiabatic losses and assuming a constant magnetic field with
time, for the JP model the cutoff energy is:
\begin{equation}
E_c {\rm[Gev]} = 1.3 \times 10^{10} \frac{1}{(B_{eq} ^2 + B_{IC} ^2)t} \; ,
\end{equation} where B$_{IC}$ represents the equivalent magnetic field strength of
the microwave background and is expressed in $\mu$G, as well as
$B_{eq}$.\\ The break frequency is
\begin{equation}
\nu^* {\rm (MHz)}= 16.08 B_{eq} E_c ^2 \; .
\end{equation}
We have fitted the integrated fluxes of the Beaver for two regions
along the tail (labeled A and B in Fig. \ref{comparisonbeaver}). They
have been selected far from any possible contaminant source, as
nucleus and extended emitting areas of the cluster radio halo. The
tail of the Beaver was not detected at 25 cm in region B, thus the
flux we used in this case is an upper limit. To produce the fit, the
break frequency and the spectral index in the part of the spectrum not
affected by any evolution ($\alpha_{inj}$) are left as free
parameters. The result is shown in Fig. \ref{jpmodels}. The very low
$\chi^2$ value may be due to the number of free parameters used in the
fit. For region A, the best fit is obtained for $\alpha_{inj} = - 0.5$
and a break frequency $\nu^* = 353.6 {\rm MHz}$. For region B, we have
assumed the same injection spectral index than region A, obtaining
$\nu^*$ = 94.52 MHz.

From the shape of its synchrotron spectrum and under the assumption of
equipartition, it is possible to estimate the radiative age of the
radio source. For the JP model, this is given by
\begin{equation}
t_r = \frac{1.6 \times 10^{19}}{B^2 + B_{IC} ^2} \sqrt{\frac{B}{(1+z) \nu^{*}}} \;\;\;\;\;\; {\rm [yr]} \;,
\end{equation} 
where $B$ is in $\mu$G and $\nu$ in GHz. Assuming a CMB temperature
of 2.725 K, $B_{IC}$ = 3.238$(1+z)^2$ $\mu$G. For regions A and B, we
obtain the radiative ages $t_{r_{a}}$ $\sim$ 1.9 10$^8$ yr and
$t_{r_{b}}$ $\sim$ 2.2 10$^8$ yr, respectively.  For this calculation,
we assumed $B$ to be the equipartition magnetic field. $B_{eq}$ is
reported in Table \ref{physicalparametersbeavertable}, together with
the other physical parameters for regions A and B, and the head of the
Beaver. The equipartition parameters were computed using the fluxes at
85 cm (column 4). We shaped the head of the source with a spherical
geometry with a diameter of 15 kpc, while for the tail we assumed a
double cylinder geometry, with each cylinder having a depth of 25 kpc
and a length of 360 kpc. The sizes of the structures were derived from
the high resolution ($2{\arcsec}\times2{\arcsec}$) image of the Beaver
in Fig. 1 in \citet{2006A&A...460..425G}.The depth and the width are
assumed to be the same for each feature.  To avoid any overestimation
of the physical parameters, we adopted a spectral index which is the
average of $\alpha_{85 cm} ^{2 m}$ and $\alpha = -0.7$. For position
B, we assumed a spectral index $\alpha = -0.5$, which is the injection
spectral index as for position A.

An estimation of the kinematic age of the Beaver can be obtained
assuming that the host galaxy traveled from position B till its
current location (D = 900 kpc) with the constant velocity of $\sim$
1000 km/s (the velocity dispersion of A2255 is $\sigma_{{\rm
v}_{A2255}}$ = 1266 km/s). This implies t$_{\rm kin}$ $\sim$ 6
$\times$ 10$^8$ yr.

\vspace{2.0cm}

\begin{table*}
\caption{Fluxes and equipartition parameters of the Beaver radio galaxy.}
\label{physicalparametersbeavertable}
\smallskip
\begin{center}
\begin{minipage}{\textwidth}
\centering
{\small
\begin{tabular}{cccccccccc}
\hline
\hline
\noalign{\smallskip}
Region    & Size           & $S_{2 m}$            &  $S_{85 cm}$         & $S_{25 cm}$          & $\alpha_{85 cm} ^{2 m}$ &  $\alpha_{25 cm} ^{85 cm}$   &        ${u_{min}} ^a$         & ${B_{eq}} ^b$   &  ${P_{eq}} ^c$       \\
          & (kpc)          &    (mJy)             &     (mJy)            &    (mJy)             &                         &                              &      (erg/cm$^{-3})$      & ($\mu$G)   & (dyne cm$^{-2}$)\\
\noalign{\smallskip}	   		       	   	       		   		       														   	   
\hline			   		       		       				       								   	   						   
\noalign{\smallskip}
Head     &  15 $\times$ 15 &  472 $\pm$ 13        &  306 $\pm$ 4         &   145 $\pm$ 2        &  $- 0.5 \pm 0.1 $       &   $-0.6 \pm 0.1$             &    8.1 10$^{-11}$         &   30       &  2.7 10$^{-11}$  \\
\noalign{\smallskip}	   		       	   	       		   		       														   	   
\hline			   		       		       				       								   	   						   
\noalign{\smallskip}
A        & 2 $^d$ $\times$ (25 $\times$ 360) &  113 $\pm$ 6         &   35 $\pm$ 1         &   1.2 $\pm$ 0.2      &   $-1.4 \pm 0.1$        &   $-2.7 \pm 1.1$             &    1.9 10$^{-13}$         &   1.4      & 6.4 10$^{-14}$  \\ 
\noalign{\smallskip}	   		       						       																	   
\hline			   		       						       														   	
\noalign{\smallskip}	   		       						       																	   
B        & 2 $^d$ $\times$ (25 $\times$ 360) &  99  $\pm$ 6         &   6  $\pm$ 1         &   $<$ 1.1            &   $-3.3 \pm 0.2$        &   $< - 1.4$                  &    1.6  10$^{-14}$        &   0.4      & 5.6 10$^{-15}$  \\ 
\noalign{\smallskip}
\hline
\noalign{\smallskip}
\noalign{\smallskip}
\multicolumn{9}{l}{$^a$ Minimum energy density}\\
\multicolumn{9}{l}{$^b$ Equipartition magnetic field}\\
\multicolumn{9}{l}{$^c$ Equipartition pressure}\\
\multicolumn{9}{l}{$^d$ The tail is shaped as a double cylinder}
\end{tabular}
}
\end{minipage}
\end{center}
\end{table*}

\begin{table}
\caption{Physical parameters of the X-ray emitting gas at the location of the Beaver.}
\label{staticpressure}
\smallskip
\begin{center}
{\small
\begin{tabular}{cccc}
\hline
\hline
\noalign{\smallskip}
Region &    $r$ $^a$   &    $n(r)$ $^b$       &   $P_{st}$ $^c$     \\
       &  (kpc)  &   cm$^{-3}$     &  (dyne cm$^{-2}$)  \\
\noalign{\smallskip}	 	   	       			       	   	       		   			   
\hline					       			       		       				   	   						   
\noalign{\smallskip}
Head   &   1.5 10$^3$&  1 10$^{-4}$    &  1.2 10$^{-12}$    \\
\noalign{\smallskip}	 	   	       			       	   	       		   			   
\hline					       			       		       				   	   						   
\noalign{\smallskip}
A      &   1.1 10$^3$ &  2.2 10$^{-4}$  &  2.5 10$^{-12}$    \\ 
\noalign{\smallskip}	 	   	       			       	   	       		   			   
\hline					       			       		       				   	   						   
\noalign{\smallskip}
B      &   7.8 10$^2$&  4.1 10$^{-4}$  &  4.6 10$^{-12}$    \\ 
\noalign{\smallskip}
\hline
\noalign{\smallskip}
\noalign{\smallskip}
\multicolumn{4}{l}{$^a$ Projected distance from the X-ray centroid}\\
\multicolumn{4}{l}{$^b$ Electron density}\\
\multicolumn{4}{l}{$^c$ Static pressure}
\end{tabular}
}
\end{center}
\end{table}

\subsection{Radio source confinement}
\label{physicalparametersbeaver}

An important connection between the ICM and the radio galaxies is
provided by the thermal pressure of the gas.  From the radio data it
is possible to derive the pressure within the radio source, under the
assumption that the radio galaxy is in equipartition. From X-ray data,
one can estimate the parameters of the intergalactic medium
surrounding the radio source. By comparing the internal pressure of
the radio emitting plasma with the thermal pressure of the ambient
gas, we can get information about the equilibrium between thermal and
non-thermal plasma.

The X-ray emitting gas plays an important role in influencing the
expansion and the structure of a radio source because it gives rise to
a static pressure ($P_{st}$), which depends only on the gas
temperature (T$_{ig}$) and on the numerical density (n$_{ig}$) of the
X-ray emitting gas
\begin{eqnarray}
\label{static}
P_{st} = 2n_{ig} k T_{ig} \; ,
\end{eqnarray}
where $k$ is the Boltzmann constant.

For our analysis, we used the X-ray results obtained by \citet{fer},
and we scaled the parameters to our cosmology. The central electron
density of the X-ray emitting gas is $n_0$ = 2.05 10$^{-3}$ cm$^{-3}$
and its temperature is $\sim$ 3.5 keV, corresponding to $3.5 \times
10^7$ K. The thermal pressure of the X-ray emitting gas has been
scaled with the $\beta$ model law:
\begin{equation}
n(r)=\frac{n_0}{[1+(\frac{r}{r_c})^2]^{3\beta /2}} \; ,
\end{equation} 
where $r_c$ = 4.8$\hbox{$^{\prime}$}$ $\pm$ 0.4$\hbox{$^{\prime}$}$
and $\beta$ = 0.74 $\pm$ 0.04.

The Beaver radio galaxy lies at a projected distance from the center
of the cluster of $\sim$ 18$\hbox{$^{\prime}$}$ (1.6 Mpc), where the
X-ray brightness of the thermal gas is very low, but still above the
background (see Table \ref{staticpressure}). By comparing the internal
pressure of regions A and B along the tail (see Table
\ref{physicalparametersbeavertable}) with the static pressure of the
X-ray emitting gas at their projected distance from the cluster center
computed with equation \ref{static} (Table \ref{staticpressure}), we
conclude that the equipartition pressure is lower by a factor of
$10^2$ or $10^3$ than the corresponding thermal pressure of the
gas. We notice that projection effects might play an important role in
the determination of ${P}_{st}$. The large gap between
${P}_{eq}$ and ${P}_{st}$ implies that either the numerous
assumptions used in the calculations of the equipartition parameters
are not valid, or that there is a real deviation from the
equipartition conditions.\\

\section{Discussion}
\label{discussion}

\subsection{Halo origin}
\label{haloorigin}

The origin and evolution of cluster radio halos is still a matter of
debate \citep{1977ApJ...212....1J,1980ApJ...239L..93D}. The main
difficulty in explaining their nature arises from their large size
($\sim$ 1 Mpc) and the short radiative lifetime of the relativistic
electrons emitting in them. Different theoretical models have been
suggested in order to infer the mechanism of transferring energy into
the relativistic electron population and for the origin of the
relativistic electrons themselves: in-situ re-acceleration of
relativistic electrons by shock waves ({\it primary models}), particle
injection from radio galaxies, acceleration out of the thermal pool,
secondary electrons resulting from hadronic collisions of relativistic
protons with the ICM gas protons ({\it secondary models}), and
combination of these processes
\citep{2003ASPC..301..349B,2003ASPC..301..203B,2003ASPC..301..337P}.

Several properties make the halo of A2255 unique among the other known cluster
radio halos. embedded in the halo emission there is an unusually large number
(5) of tailed radio sources, which are likely supplying it with relativistic
particles \citep{fer}. Morphologically, the structure of the halo is rather
complex, as Figs \ref{25cmfig}--\ref{2mfigure} show. At 25 cm, the halo has a
rectangular shape, due to the presence of 3 bright filaments at the edges,
perpendicular to each other.  Previous VLA observations at 1.4 GHz \citep{gov}
revealed that they are strongly polarized ($\sim$ 20\%--40\%), which
represents the first detection of a polarized halo in the literature. Also at
85 cm and at 2 m the halo has a rectangular shape, with a considerable
extension towards the S and SW with respect to high-frequency (see
Figs. \ref{85cmfigure} and \ref{2mfigure}). Furthermore, in the 2 m
observations a new extended emitting region is detected to the NW of the halo
(Figs. \ref{2mfigure} and \ref{2mfiguresubtraction}). This new feature,
undetected in the more sensitive observations at 85 cm
(Fig. \ref{85cmfigure}), should have a very steep spectrum ($\alpha \leq -
2.6$).\\ As a consequence of the presence of the filaments, the surface
brightness of the halo increases from the center towards the edges and this
reflects in the the spectral index behavior within the halo (see Figs
\ref{spix2585}--\ref{spix852}). Although the low resolution of the images does
not allow us to obtain very precise trends and the presence of the radio
galaxies at the cluster center partially contaminate the determination of the
spectral index, it is evident that $\alpha$ is steeper in
the regions in between the filaments and flattens along them, where fresh
(re-) acceleration of the relativistic particles takes place. This trend, due
to the presence of the radio filaments, has not been seen before in a radio
halo; in the few clusters for which maps of the spectral index are available
(e.g. Coma C, \citet{1993ApJ...406..399G}, A665, A2163,
\citet{2004A&A...423..111F}, A3562, \citet{2005A&A...440..867G}, A2744, and
A2219, \citet{2007A&A...467..943O}) the halo radio spectrum shows variations
on small scales (as clumpy structures) and/or on large scales (radial
steepening from the center to the edges).

The singular properties of the halo of A2255 (morphology, surface
brightness, spectral index distribution, and polarization), suggest
that either this extended structure has a more complex structure than
the other known radio halos or that it is seen in a particular stage
of its evolution. Since our observations do not support a unique
explanation for the origin of the radio halo, we suggest the two
following scenarios:
\begin{itemize}
\item{ {\it the halo and the filaments are two physically distinct
structures, seen in projection as a unique feature.}\\

 X-ray, optical and radio data provide evidence that A2255 is
 currently in an active dynamical state, and other surrounding
 structures might be interacting with it. A ROSAT X-ray survey
 observation indicates that this cluster belongs to the north ecliptic
 pole super-cluster, which contains at least 21 members
 \citep{2001ApJ...553L.115M}. Signs of the interaction of A2255 with
 the other cosmic structures are expected. Indeed, many radio features
 are detected at various distances from the cluster center. These
 radio ``relics'' suggest that the cluster is currently accreting gas
 from the cosmic environment and is possibly hosting many shock waves
 deriving from the past merger activity. Relics, usually found at the
 cluster periphery, should occasionally also be detected in projection
 towards the cluster center. Thus, it is possible that the three
 filaments that ``form'' the radio halo of A2255 are shocked regions
 of plasma which are seen in projection on the central real radio
 halo. Their highly polarized flux \citep{gov} favors a location of
 these structures in the foreground instead than in the background,
 where Faraday depolarization would rapidly depolarize them. The new
 extended feature detected at 2 m to the NW of the cluster center
 could be considered as an extension of the halo and its very steep
 spectral index could be explained in the framework of the primary
 electron re-acceleration models, where a steepening of the spectral
 index from the center to the edges of the halo is expected. However,
 since radio halos in the literature are known to have regular
 morphologies, we should also consider the hypothesis that the new
 feature is not physically related with the central radio halo. In
 this case, this could be considered as the first example of Mpc-size
 diffuse structures (MDS), which lie around clusters and are
 detectable at very low frequency only.}\\

\item{ {\it the halo has an intrinsic surface brightness increasing
towards the edges and a spectral index flattening from the center to
its outermost regions.}\\

The bright and polarized filaments at the edge of the halo could
derive from the injection in the intra cluster medium of energy on
large scales, which is produced by turbulence, resulting from past
merger activity. On the theoretical point of view, the picture of the
development of turbulence in clusters seems still
uncertain. \citet{2005MNRAS.364..753D} argue that the bulk of
turbulence is injected in the core of galaxy clusters, implying a more
developed turbulence in the innermost regions, compared to the outer
parts. On the other hand, cosmological numerical simulations suggest
that turbulence is expected to be greater at increasing radial
distances from the cluster center
\citep{1998ApJ...495...80B,2003AstL...29..783S}. Moreover,
\citet{2006A&A...460..425G} showed that to reproduce the RM
distribution for three of the radio galaxies of A2255 one needs a
power spectrum of the magnetic field of the cluster steepening from
the center towards the periphery and the presence of filamentary
structures on large scale. Therefore, it is possible that the
turbulence at the center of A2255 gave rise to shocked regions at the
edges of the halo.\\ In this framework, a physical connection between
central halo and the newly detected extended feature to the NW from it
seems unlikely. Thus, the latter should be classified as the first
example of MDS.}
\end{itemize}

\subsection{Relics}
\label{relics}

Relics are associated with clusters with or without cooling core,
suggesting that they may be related to minor or off-axis mergers, as
well as major mergers. Their formation is supposed to be related to
shocks either by Fermi-I diffuse acceleration of ICM electrons
\citep{ensslin,keshet} or by adiabatic energization of the
relativistic electrons confined in bubbles of fossil radio plasma
(``ghosts''), released by a former active radio galaxy. Shocks in
clusters environments can derive from the merger of subclusters
\citep[{\it merger shocks},][]{2008SSRv..134..119B}, or can be due to
the accretion of diffuse , unprocessed (cold) matter on
gravitationally attracting nodes \citep[{\it accretion
shocks},][]{2003MNRAS.342.1009M,2003ApJ...585..128K}

The presence of shocked regions in A2255 is strongly supported by
X-ray and radio observations. Recently, the XMM-Newton satellite
detected a shocked region near ($\sim$ 4$^{\prime}$, 360 kpc) the cluster
center . The morphology of this thermal emission suggests that it
could arise from a merger that happened along the east-west direction
about 0.15 Gyr ago \citep{sakelliou}. However, the complicated
structure of the temperature map does not allow any precise conclusion
about the geometry of the merger.

At radio wavelengths, A2255 shows the presence of several relics and
filamentary features (see Fig. \ref{25cmfig} and
Fig. \ref{85cmfigure}) which are likely associated with shocks. The
young merger event inferred by the X-ray observations cannot be
responsible for the formation of these features, which are located at
large distance from the cluster center. The shock derived from it lies
still inside the radio halo.  Cosmological simulations show that
cosmic structures form through several merger events. Consequently,
the cluster environment can host many shocks traveling towards
different directions, together with possible other shocks, appearing
at large distance from the cluster center and possibly associated with
the flow of cold gas into the potential wall of the cluster
itself. This scenario seems to be present in A2255. In its outermost
regions there are relic-like structures, which are probably associated
with LSS shocks \citep{2008A&A...481L..91P}. In addition, in its
innermost regions there are shock-related features (NE relic, {\it
bridge, C1}, and {\it C2} in Fig. \ref{25cmfig}) which likely derive
from the past merger activity.

The shape and location of the NE relic clearly suggest that this could
originate in a shock wave traveling along the NE to SW direction. The
orientation of the magnetic field in it, being parallel to the major
axis \citep{gov}, also supports this scenario.  Its radio spectral
index ($\alpha$ $\sim$ --0.7), unusually flat for a relic, suggests
that this structure is young and the relativistic electrons have been
recently (re)accelerated.  The spectral index trend along the minor
axis, being steep near the cluster center and flattening towards the
periphery (see left panel of Fig. \ref{spixprofileNERELIC}), implies
that the shock is traveling outwards. The fact that we see a gradient
of the spectral index also along the major axis (see
right panel of Fig. \ref{spixprofileNERELIC}) , may be due to the
complex geometry of the shock. The presence of filamentary radio
emission perpendicular to the NE relic ({\it bridge, C1} and {\it C2}
in Fig. \ref{25cmfig}) suggests that another shock could be present at
this location and it could have influenced the physical properties of
the NE relic itself.

The nature of the NW and SW relics has been already investigated by
\citet{2008A&A...481L..91P} by means of the 85 cm observations.  The
spectral index study carried out in this paper, allow us to reinforce
our first conclusion. The spectrum of the NW relic has been determined
for the only filament oriented towards the cluster center. Its
$\alpha$ values (--1.2 $\pm$ 0.08 $\le \alpha \le$ --0.9 $\pm$ 0.1)
and its location with respect to the cluster center, still support the
hypothesis that it could be related to LSS shocks. The SW relic is not
detected at 2 m, therefore we can only compute a lower limit ( $\alpha
\geq -1.2$).  Given the common physical properties with the NW relic,
we still suggest that also the SW relic could be associated with LSS
shocks.

\subsection{The Beaver radio galaxy}
\label{beaverdiscussion}

Narrow angle tail (NAT) radio galaxies have a U shape morphology, with
the nucleus coincident with the parent optical galaxy and both tails
bending backwards. Most of these sources were detected in clusters of
galaxies and they have been the target of numerous investigations. The
morphology of NAT radio sources is due to the ram pressure exerted by
the ambient medium on the radio plasma ejected by the host galaxy that
moves through the cluster. In this scenario, the quasi continuous
beams ejected by the central galaxy are bent backwards, forming a tail
which marks the path that the galaxy has traveled through the ICM. It
is currently accepted that in non collapsing clusters, the motion of
the ICM is subsonic, thus not having a substantial effect on the
orientation of the radio tails. In this context, the tail of the radio
galaxy will be parallel to the galaxy advance motion.\\ The size of
NAT radio galaxies, from the nucleus till the end of the tail, is on
average about 200--300 kpc \citep{1987AJ.....94....1V}. This is
comparable with the extent of the more common double radio sources,
which have a I shape, with the parent galaxy located at the
centroid. However, a few examples of NAT radio galaxies with
uncommonly long tails are reported in the literature. Among them, the
most well known objects are IC 711 in Abell 1314 ($\sim$ 650 kpc,
\citet{1987AJ.....94....1V,1988Ap&SS.149..225V}) and NGC 1265 in the
Perseus cluster ($>$ 500 kpc, \citet{1998A&A...331..901S}). The Beaver
radio galaxy (1712+638, J2000), whose tail extends to almost 1 Mpc at
85 cm wavelength, provides a clear example of such radio sources.

The long tail of the Beaver may suggest that the host galaxy (2MASX
J17131603+6347378) has traveled from the central regions of the
cluster to its actual location without any substantial deviation from
its original direction, at least on the plane of the sky. Projection
effects could make the tail appear shorter, but knowing the real
extent of the tail as well as the 3-dimensional location of the Beaver
with respect to the other cluster structures is challenging. A
polarization study of the cluster at 85 cm reveals that the end of the
tail of the Beaver is polarized (Pizzo et al., {\it in prep.}). This
suggests that this structure is not located deep inside the dense
X-ray emitting gas, which would make it depolarized due to Faraday
depolarization. However, since the radial velocity of the galaxy
(v$_{Beaver}$ = 24843 km/s) is only 700 km/s higher than the mean
velocity of the cluster (v$_{A2255}$ = 24163 km/s), which is less than
its velocity dispersion ($\sigma_{{\rm v}_{A2255}}$ = 1266 km/s), the
direction of the velocity vector cannot be determined.

We have checked that the long nature of the Beaver is real subtracting
the point sources visible at higher frequency along the path of the
tail in the low-frequency maps. Support to the genuine nature of the
tail is also the regular steepening of the spectral index from the
head towards its outermost regions (Figs. \ref{beaverspix2585} and
\ref{beaverspix852}).

The severe steepening of the spectral index along the tail of the
Beaver implies that the relativistic electrons responsible of the
radio emission suffered important energy losses after their first
ejection from the parent galaxy, which is confirmed by the high
radiative ages of the plasma at the end of the tail ($\sim$ 2.2 1$0^8$ yr, see Sect. \ref{spectralageing}). Its detectability, even
at low frequency, raises questions, since adiabatic expansion should
play an important role in rapidly depriving the electrons of their
energy. In this context, it is suggested that the ICM could have a
drastic influence on the final stages of the life of the plasma
\citep[e.g.][]{1998MNRAS.298.1113V,2007A&A...470..875P}. In our case,
the high static pressure exerted by the ICM on the ending part of the
tail of the Beaver (where the internal pressure is $10^3$ times lower
than the external one) may prevent its quick dissipation through
adiabatic expansion.

It is worth noting that the Beaver shows at the end of the tail
spectral index values very similar to the ones of the southern regions
of the halo (Fig. \ref{beaverspix852}). This characteristic suggests
that the Beaver and the halo may be physically related structures and
that the radio galaxy provides the halo with the relativistic
particles for its own radio emission. Tailed radio sources are thought
to supply the relativistic electrons to radio halos \citep[see the
example of NGC4869 in the COMA cluster,][]{1993ApJ...406..399G}. In
the Beaver radio galaxy and A2255 we may witness this process.

\section{Summary and conclusions}
\label{summary}

We presented WSRT observations of the cluster of galaxies A2255 at
wavelengths of 25 cm, 85 cm, and 2 m. In the final maps, the radio halo
and the relics are detected. In each image, the radio emission
associated with the cluster seems to be very complex and several new
features are detected. Our observations, together with the results
coming from optical and X-ray studies, have highlighted an interesting
picture for this cluster, which has probably undergone multiple merger
events during its past history. Several features, detected in the
radio domain in and around the cluster, could be considered as an
indication of a still undergoing strong dynamical activity.

 At 25 cm, the radio halo shows a U shape, with three bright filaments
perpendicular to each other at the edges. Two additional filamentary
features are detected at low level, the first one at the same location
of the NW relic found at 85 cm and the other one near the already
known NE relic (see Fig. \ref{25cmfig}).
 At 85 cm, the halo is more extended towards S and SW and at 2 m it
 shows a Mpc-size extension towards NW which was not detected
 previously.

From the spectral index images we found that $\alpha$ is steep in the
 central part of the halo and flattens moving towards its outermost
 regions (Figs. \ref{spix2585} and \ref{spix852}).  This is likely due
 to the presence at this location of bright filaments, clearly
 detected at higher frequency.  Understanding the nature of these
 filaments and of the newly detected extended emission between halo
 and NW relic is challenging. One possibility is that the central
 radio halo and the filaments are two physically unrelated structures,
 seen in projection near the cluster center. In this case, the
 extended feature detected at 2 m to the NW of the halo could be
 considered as an asymmetric extension of the halo itself; its steep
 spectrum ($\alpha \leq - 2.6$) could be justified in the framework of
 the primary electron re-acceleration models, where a steepening of
 the spectral index from the center to the edges of the halo is
 expected. However, since radio halos are known in the literature as
 structures showing a regular morphology, the new feature could be
 physically not related with the central radio halo and it should be
 considered as the first example of Mpc-size diffuse structures
 (MDS), which are detectable around clusters at very low frequency
 only.  On the other hand, it is also possible that A2255 hosts an
 intrinsically peculiar radio halo, which has a filamentary structure
 at the edges. In this scenario, the diffuse emission region to the NW
 of the halo should be considered as not related to the central halo
 and classified as MDS.  In order to distinguish between the two
 possible scenarios, it is important to understand the real nature of
 the filaments. Sensitive X-ray observations are needed to investigate
 the presence of X-ray substructures, possible shocks, and their
 connection to the radio halo. On the other hand, a detailed study of
 the rotation measure of the different physical structures of the
 cluster could make it possible to infer the 3-dimensional geometry of
 A2255 (Pizzo et al., {\it in prep.}), possibly answering this still
 open question.

The NE relic shows a flattening of the spectral index along its minor
axis, moving outwards from the cluster center (see left panel of
Fig. \ref{spixprofileNERELIC}). This, and the rather flat integrated
spectrum of the NE relic (Fig. \ref{integratedspectrahalo&relic},
right panel), suggests that the relativistic particles have been
recently (re) accelerated by a shock, that is traveling from the
center of the cluster towards its periphery along the NE-SW direction.

The NW and SW relics are both detected at 85 cm only. The NW relic is
also visible at 2 m, but just partially at 25 cm. From the spectral
index image between 2 m and 85 cm, we obtain spectral index values
ranging between --0.6 and --1.8, steepening towards the cluster
center. We cannot obtain a direct estimation of the spectral index for
the SW relic, because of sensitivity limitations in the 25 cm and 2 m
maps, therefore only a lower limit can be derived in this case
($\alpha$ $\geq$ --1.2).

The tail of the Beaver radio galaxy increases its length to almost 1
Mpc between 25 cm and 85 cm. This is a clear indication that the
plasma along the tail suffered severe energy losses after the ejection
from the parent optical galaxy, and that is still confined in the tail
due to the static pressure exerted by the external ICM. The very long
tail of the radio galaxy gives us the possibility to test ageing
models for the relativistic electrons. The JP model gives a good
representation of our data for the initial part of the tail
(Fig. \ref{jpmodels}, left panel). For the ending part, more
observations at intermediate wavelengths between 25 cm and 85 cm are
needed to draw any conclusion (Fig. \ref{jpmodels}, right panel).  In
the spectral index map between 85 cm and 2 m
(Fig. \ref{beaverspix852}) there is the indication that the $\alpha$
values of the ending part of the tail of the Beaver are similar to the
ones of the southern regions of the halo of A2255. This might suggest
that the radio galaxy has provided the halo with relativistic
particles for its own radio emission. To test this hypothesis, a more
detailed study of the steepening of the spectral index along the tail
of the Beaver is needed. LOFAR, thanks to its wide low-frequency range
and high resolution, will play a major role in such
investigations.

\begin{acknowledgements}
R.F.P. is thankful to Luigina Feretti for the useful discussions
during the data analysis. R.F.P. is grateful to Monica Orienti for the
very helpful and detailed comments and suggestions during the writing
of the manuscript.  The Westerbork Synthesis Radio Telescope is
operated by ASTRON (Netherlands Institute for Radio Astronomy) with
support from the Netherlands Foundation for Scientific Research (NWO).
\end{acknowledgements}

\bibliographystyle{aa}
\bibliography{12465col}
\end{document}